\documentclass[11pt,english,dvipsnames]{article}
\usepackage[utf8]{inputenc}
\usepackage[english]{babel}
\usepackage{booktabs,float}
\usepackage{amsmath,amssymb,amsthm,amsfonts}
\usepackage{enumerate}
\usepackage{tikz}
\usepackage{xcolor}
\usepackage[authoryear,round]{natbib}
\usepackage[a4paper,body={16cm,23.7cm}]{geometry}
\usetikzlibrary{positioning,decorations.pathreplacing}

\newtheorem{theorem}{Theorem}
\newtheorem{lemma}{Lemma}

\newtheorem{remark}{Remark}
\newtheorem{example}{Example}

\begin{document}

\title{\textbf{MEASURES OF RELEVANCE TO THE SUCCESS OF STREAMING PLATFORMS}\thanks{Juan Carlos Gon\c{c}alves-Dosantos acknowledges the grant PID2021-12403030NB-C31 funded by MCIN/AEI/10.13039/501100011033 and by ``ERDF A way of making Europe/EU''. He also acknowledges financial support from Xunta de Galicia (Grupos de Referencia Competitiva ED431C-2020/14) and from CITIC that is supported by Xunta de Galicia (convenio de colaboraci\'on entre la Conseller\'ia de Cultura, Educaci\'on, Formaci\'on Profesional e Universidades y las universidades gallegas para el refuerzo de los centros de investigaci\'n del Sistema Universitario de Galicia (CIGUS)). Ricardo Mart\'inez acknowledges the R\&D\&I project grant PID2020-114309GB-I00 funded by MCIN AEI/10.13039/501100011033 and by ``ERDF A way of making Europe/EU'', and he also acknowledges financial support from Junta de Andaluc\'ia under projects FEDER UGR-A-SEJ-14-UGR20 and Grupos PAIDI SEJ660. Joaqu\'in S\'anchez-Soriano acknowledges financial support from the Generalitat Valenciana under project PROMETEO/2021/063.}}
\author{Juan Carlos Gon\c{c}alves-Dosantos \\ Universidad Miguel Hern\'andez de Elche \and Ricardo Mart\'inez \\ Universidad de Granada \and Joaqu\'in S\'anchez-Soriano \\ Universidad Miguel Hern\'andez de Elche}
\date{}
\maketitle
\begin{abstract}
Digital streaming platforms, including Twitch, Spotify, Netflix, Disney, and Kindle, have emerged as one of the main sources of entertainment with significant growth potential. Many of these platforms distribute royalties among streamers, artists, producers, or writers based on their impact. In this paper, we measure the relevance of each of these contributors to the overall success of the platform, which is information that can play a key role in revenue allocation. We perform an axiomatic analysis to provide normative foundations for three relevance metrics: the uniform, the proportional, and the subscriber-proportional indicators. The last two indicators implement the so-called pro-rata and user-centric models, which are extensively applied to distribute revenues in the music streaming market. The axioms we propose formalize different principles of fairness, stability, and non-manipulability, and are tailor-made for the streaming context. We complete our analysis with a case study that measures the influence of the 19 most-followed streamers worldwide on the Twitch platform.
\end{abstract}
\textbf{Keywords}: Measure, relevance, proportionality, streaming, axiom

\newpage

\section{Introduction}

In 2022, Netflix, Amazon Prime Video, Disney+, and HBO MAX -- four of the most successful streaming platforms -- reported a joint annual revenue of \$103,5 billion. These companies operate so-called over-the-top broadcasting services by offering to their customers a media library of movies, series, and documentaries. Viewers must pay a monthly fee to obtain unlimited access to the catalog. As Table \ref{Table_revenues} shows, the subscription prices for these platforms are similar, but the number of subscribers and earned revenue differ significantly. 

\begin{table}[H]
\centering
\begin{tabular}{lrrr}
\toprule    
    Company & Subscription price & Subscribers & Revenue \\
\midrule
    Netflix & \$7 & 231 million & \$31.6 billion \\
    Amazon Prime Video & \$9 & 200 million & \$25.2 billion \\
    Disney+ & \$8 & 164 million & \$7.4 billion \\
    HBO+ & \$10 & 95 million & \$39.3 billion \\
\bottomrule
\end{tabular}
\caption{Fees, subscribers, and revenues of the main streaming platforms.\label{Table_revenues}}
\end{table}

Two elements are crucial for determining the success (or revenue) of a platform: the fee and the catalog. Some companies have only two or three shows each year that subscribers are interested in, but those shows may be massive hits. Other platforms choose a different model, offering a wider catalog based on the variety and quantity of content. Be that as it may, it is clear that not all titles offered by these companies are equally in demand, and the available titles may be decisive for consumers to decide which platform they subscribe to. Therefore, if one platform offers content that is more relevant than others, a worthwhile question to ask is: what is the impact of each single show on the overall success of a platform? This question can be answered by applying the theoretical model presented in this paper.

The previous setting is not unique to over-the-top subscription video-on-demand providers; it can be extended to many similar situations, in which selling products in a package may be more profitable than selling them independently (see \cite{Adams76}). Spotify, Twitch, YouTube, and other web-based services are only a few examples. In all these, subscribers pay a subscription fee to obtain unlimited access to a set of services, and it is only natural to wonder what is the individual contribution of each of those services to the success of the whole package.

In our model, a \emph{platform} is described by four elements: the set of \emph{services} the platform provides (TV shows, artists, streamers, books, etc.), the set of \emph{subscribers} who have unlimited access to those services (viewers, users, readers, etc.), the \emph{subscription price} paid by each subscriber, and the \emph{consumption matrix} that indicates the quantity of each service consumed by each subscriber.\footnote{Our setting is general enough to allow the subscription price to vary across subscribers. Fixing a common subscription fee is possible, but it is just a particular case in our model.} An \emph{indicator} is a measure that determines the relevance of each single service to the success of the platform, which is the total revenue generated from selling subscriptions. 

In this study, we introduce four natural indicators to determine the share of the platform's success due to each service. The \emph{uniform indicator} stipulates that all services are equally relevant. The \emph{subscriber-uniform indicator} states that the relevance of each individual subscriber is divided uniformly among the services that the subscriber consumes. The \emph{proportional indicator} assigns, to each service, a relevance proportional to its aggregate consumption. The last proposal, the \emph{subscriber-proportional indicator}, also considers a proportional perspective, but applied to each single subscriber. More precisely, the success of each single subscriber is proportionally assigned among the services that the subscriber consumes. The relevance of each service is the lump sum of those proportions across subscribers. In the music streaming industry, the two most relevant remuneration methods are the \emph{pro-rate} and the \emph{user-centric} mechanisms. The former is based on an aggregate approach, where the overall revenue from all music subscriptions is distributed among the artists proportionally to their total number of streams. In the latter, the calculation is based on the listening habits of each individual subscriber; the subscription fee she pays is proportionally distributed, but only among the artists whose music this subscriber consumes. The \emph{proportional indicator} and the \emph{subscriber-proportional indicator} implement the \emph{pro-rate} and \emph{user-centric} principles, respectively.

As in \cite{Manshadi23} and \cite{Singal22}, we follow the axiomatic methodology to investigate the existence of indicators that satisfy combinations of properties (called \emph{axioms}) that are suitable for our setting. The axioms that we propose formalize different principles related to fairness, stability, and absence of manipulability that an indicator of relevance should satisfy. The first property is \emph{efficiency}, which requires that the sum of the relevance of all services is equal to the total relevance of the platform. Regarding fairness, \emph{symmetry} requires that if two services are equal (i.e., they are equally consumed by all the subscribers), then they must be equally relevant to the success of the platform. \emph{Strong symmetry} requires that if two services are consumed by the same subscribers, then both must be equally relevant. \emph{Nullity service} states that services not consumed have no relevance to the platform. With regard to stability, \emph{homogeneity} states that the relevance must not be affected by the units (thousands or millions, for example) in which the consumption matrix is expressed. \emph{Consumption sensitivity} requires that small mistakes in the measurement of the consumption do not drastically alter the significance of each service. \emph{Composition} states that when data is combined from two subgroups, it is possible to determine the indicator of the total group by a suitable composition of the subgroups' indicators. \emph{Sharing proofness} and \emph{non-manipulability} belong to the third type of principles. The former states that if subscribers share their subscriptions with friends and relatives, then the relevance of the services remains unaltered. The latter axiom requires that the indicator should not be able to be manipulated by merging or splitting services.

Regarding our findings, we obtain some characterizations of the aforementioned indicators. More precisely, if we require both \emph{composition} and \emph{non-manipulability} to be satisfied, we must apply the \emph{subscriber-proportional indicator}. Theorem \ref{thm2} shows that the \emph{proportional indicator} is the unique indicator that fulfills \emph{sharing proofness} and \emph{non-manipulability} together. In Theorem \ref{thm3}, we prove that the combination of \emph{symmetry}, \emph{homogeneity}, and \emph{consumption sensitivity} unambiguously yield to the \emph{uniform indicator}. Finally, Theorem \ref{thm4} shows that if \emph{strong symmetry}, \emph{nullity}, and \emph{composition} are demanded, then the \emph{subscriber-uniform indicator} should be implemented. As in our setting, the \emph{proportional} and the \emph{subscriber-proportional} reflect the \emph{pro-rate} and \emph{user-centric} schemes. Theorems \ref{thm1} and \ref{thm2} also reveal characterizations of these two mechanisms. In other words, based on several principles of fairness, stability, and non-manipulability, we provide normative foundations for the two most-used revenue allocation methods in the music streaming industry. On the other hand, all referred indicators are examined from a game-theoretical perspective. Specifically, we demonstrate that these indicators coincide with the Shapley value (\cite{Shapley53}), which is a prominent solution concept in game theory (\cite{Roth88}, \cite{Algaba19}), of some cooperative games that can be defined for the class of problems analyzed in this paper, which reflect some aspect of how relevance is obtained. Likewise, an analysis of the indicators and properties is performed in the particular context of streaming platforms. Next, in the context of streaming, different aspects related to the information used, the structure of the indicators, and the properties they satisfy are studied. Moreover, the impact on relevance measures is analyzed when changes occur in the visualization matrix and, in this sense, which indicators streamers may prefer. Likewise, we also analyze the measures of relevance that may be the most interesting for a platform according to its own business.

We conclude our paper with an illustrative application of the indicators that we characterize. In particular, we measure the relevance of each of the top 19 most-watched Twitch streamers. This information could play a key role in determining the distribution of revenues. We highlight and explain the differences and implications of the four indicators that we analyze in this work. Based on the obtained results, we conclude that the mere use of either the number of viewers or exclusive viewers, which is the usual standard in this industry, is not the most effective approach for determining the impact of each streamer on the platform.

\subsection{Related literature}

Our work provides a novel approach to determining and valuing the impact of each single item, service, or contributor to the collective success of a platform or bundling product. Other papers have also addressed the question of how to isolate and measure the merit of single agents in the collective success of online frameworks. \cite{Singal22} provide an axiomatic justification of the \emph{counterfactual adjusted Shapley value}, which measures the contribution of individual advertiser actions (emails, display ads, search ads, etc.) to eventual customer acquisition. The authors show that this Shapley-based metric coincides with an \emph{adjusted unique-uniform} attribution scheme. In contrast, the indicators that we characterize in this work are based on proportionality rather than uniformity. Proportional schemes are usually perceived by society as fair allocation methods and, therefore, as less controversial. 

The discussion around how to divide the revenue obtained by subscriptions to music streaming platforms among artists has become particularly relevant in recent years. In this context, the \emph{pro-rata} and the \emph{user-centric} schemes have emerged as the most prominent and applied methods (see, for example, \cite{Dimont17}, \cite{Page18}, \cite{Page18b}, \cite{Prorata18}, and \cite{Meyn23}). \cite{Alaei22} study the strategic implications of these two schemes in a model with a two-sided streaming service platform that generates revenues by charging users a subscription fee for unlimited access to the content. We contribute to this discussion by providing normative foundations of the \emph{pro-rata} and \emph{user-centric} rules.

\cite{Lopez19} study the problem of how to allocate the revenues generated in a Smart TV ecosystem between the provider and the content producers. They characterize the core of the game associated with the situation and provide simple formulas for their Shapley and Tijs values, which belong to the core of the game. \cite{Lopez23} study how to allocate the revenues generated in a video platform such as YouTube, considering the navigation of the users in the platform. For this, they define dynamic games associated with the problem and provide several allocation schemes based on the structure of the Shapley value. Our approach differs from that of \cite{Lopez19,Lopez23} since they approach the matter from a game-theoretical perspective and, in our case, from an axiomatic perspective. \cite{Bergantinos23} also study the problem of how to share revenue in the streaming industry, with a particular focus on music streaming services such as Spotify. They develop and characterize allocation mechanisms based on the \emph{pro-rata} and the \emph{user-centric} principles, among others. The axioms we propose in this paper differ from those presented in \cite{Bergantinos23}. In particular, the results obtained by these authors require properties such as \emph{proportionality on streams}, \emph{equal importance of similar users}, or \emph{non-manipulability by leaving the platform}, which we do not explore in this work. The definition of indicator is also different, as we require \emph{efficiency} while they do not.

A related paper to this, but from the perspective of claim problems and generalizations, is \cite{Ju07}. The authors describe a general resource allocation problem where an infinitely divisible good must be distributed among a set of entities. Each entity is characterized by a vector, with each value reflecting the entity's relevance to a specific set of issues. In contrast to the problem discussed in \cite{Ju07}, where one infinitely divisible good is allocated, in our model we consider a vector of infinitely divisible goods, one for each subscriber, the issues in their problem. Where it depends on the subscription fee of each individual subscriber. In addition to the proportional rules explored in \cite{Ju07}, our approach allows us to consider rules that are applied to each subscriber independently. Based on the normative framework proposed by \cite{Ju07}, which encompasses a wide range of resource allocation problems, we will introduce several axioms that are relevant to the context at hand.

Other authors have addressed similar questions from different perspectives in the literature on allocation and attribution problems. \cite{Ginsburgh03}, \cite{Bergantinos15} and \cite{Martinez23} analyze the so-called museum pass problem to distribute the revenue obtained by selling museum passes that allow entrance into a set of museums. \cite{Martinez21} develop a theoretical model to measure the relative relevance of different pathologies to the lethality of a disease in a society. Additionally, in the context of biology applications, \cite{Moretti07}, \cite{Albino08}, and \cite{Lucchetti10} propose indices to identify the genes that are useful for the diagnosis and prognosis of specific diseases and cancers, while \cite{Martinez22d} report that the weighted averages of the \emph{incidence rate}, \emph{morbidity rate}, and \emph{mortality rate} are the appropriate methods to evaluate the impact of pandemics. \cite{Brander11} offer an account of the importance of the attribution problem in climate change, and \cite{Burger20} conduct an in-depth review of the attribution problem in the context of climate change, both from a technical perspective and its legal and policy applications. \cite{Manshadi23} analyze how governmental and non-profit organizations must face the task of obtaining equitable and efficient rationing of a social good among individuals whose needs (as was the case in the recent pandemic) emerge sequentially and are possibly correlated. Finally, \cite{Martinez22b} characterize several methods to evaluate the multidimensional proposals of bidders in public procurement tenders. However, our approach differs from the above mentioned papers in the problem we study, in the structure of the elements that define the problem and in several of the properties used to characterize the solutions presented.

The rest of the paper is organized as follows. In Section 2, we present the model and introduce four indicators to measure the relevance of services in a platform. In Section 3, we propose and formalize the normative framework we consider. We present the normative analysis of the indicators in Section 4, and we analyze the indicators in the streaming context in Section 5. In Section 6, we provide an application of our theoretical model. Concluding remarks are provided in Section 7.

\section{Mathematical model and relevance indicators}\label{model}

Let $\mathbb{N}$ represent the set of natural numbers, and let $\mathcal{N}$ be the set of all finite and non-empty subsets of $\mathbb{N}$. A \textbf{platform} is described by a 4-tuple $(N,S,p,C)$, where $N = \{1,\ldots, |N|\} \in \mathcal{N}$ ($|N| \geq 3$)\footnote{Just as \cite{Ju07} requires this condition for its results, we also require it for the results presented here. However, it is not a limiting condition, as the number of services is typically greater than or equal to 3 on most platforms.} is the set of \textbf{services} provided by the platform, $S =\{1,\ldots,|S|\} \in \mathcal{N}$ is the set of \textbf{subscribers} who have unlimited access to the services in $N$, $p= (p_1,\ldots,p_{|S|}) \in \mathbb{R}^{|S|}_{++}$ represent the \textbf{subscription prices} paid by each subscriber, and $C$ is the \textbf{consumption matrix}, each of whose entries $C_{is} \in \mathbb{R}_+$ indicates the quantity of a service $i$ consumed by subscriber $s$. The \textbf{success} of the platform is the total revenue generated from selling subscriptions, $\sigma=\|p\|$.\footnote{The notation $\|\cdot\|$ refers to the sum of all absolute values of all entries of a vector or matrix.} We denote by $\mathcal{D}$ the set of all platforms.

We denote by $C_{i \cdot}$ the $i$-th row of $C$, which represents the consumption of service $i$. We also denote by $C_{\cdot s}$ the $s$-th column of $C$, which represents the consumption of subscriber $s$. Therefore, $\|C_{i \cdot}\| = \sum_{s \in S} C_{is}$ and $\|C_{\cdot s}\| = \sum_{i \in N} C_{is}$ are the total consumption of service $i \in N$ and subscriber $s \in S$, respectively. We restrict ourselves to consumption matrices such that $\|C_{\cdot s}\|>0$ for all $s \in S$, that is, we assume that every subscriber has paid the subscription fee with the willingness to consume part of the content that the platform offers. Given a set of services $N' \subseteq N $, or a set of subscribers $S' \subseteq S$, the matrices resulting from removing the rows in $N'$ and the columns in $S'$ are denoted by $C_{N \backslash N'}$ and $C_{S \backslash S'}$, respectively.

The relevance of each service to the success of the platform is measured using an \textbf{indicator} $R$, which is a mapping $R: \mathcal{D} \longrightarrow \mathbb{R}^{|N|}_+$ such that
$$
\sum_{i \in N} R_i(N,S,p,C)=\sigma.
$$
For each $i \in N$, $R_i(N,S,p,C)$ indicates the share of the success of the platform that is due to $i$.

Now, we present some examples of possible indicators. The first one is straightforward; it attributes equal responsibility to all services.

\textbf{Uniform indicator}. For each $(N,S,p,C) \in \mathcal{D}$ and each $i \in N$,
$$
R^U_i (N,S,p,C) = \frac{\sigma}{|N|}.
$$

The second indicator uniformly distributes the success of each subscriber among the services that have been consumed, regardless of the consumption intensity. If subscriber $s\in S$ has not used the service supplied by $i\in N$, that is, if $C_{is}=0$, then this has no influence on subscriber $s$.

\textbf{Subscriber-uniform indicator}. For each $(N,S,p,C) \in \mathcal{D}$ and each $i \in N$,
$$
R^{SU}_i(N,S,p,C) = \sum_{s \in S, i\in N_s} \dfrac{1}{\left| N_s\right|}p_s,
$$
where $ N_s=\left\{ j \in N: C_{js} \neq 0\right\}$.

The next indicator is also natural since it divides the success proportionally to the consumption of each service. It implements the so-called \emph{pro-rata rule}, which is one of the two most prominent remuneration methods in the literature on the music streaming industry.

\textbf{Proportional indicator}. For each $(N,S,p,C) \in \mathcal{D}$ and each $i \in N$,
$$
R^P_i(N,S,p,C) = \frac{\|C_{i \cdot}\|}{\displaystyle\sum_{j \in N} \|C_{j \cdot}\|} \sigma.
$$

While the previous indicator distributes the success as a whole, the following measure is based on determining the role of each service in the success of attracting the interest of each single subscriber. Consider a subscriber $s \in S$ who has paid the subscription fee $p_s$. At this individual level, the success of the platform is simply $p_s$. Now, divide $p_s$ among all the services consumed by $s$ in proportion to the consumption; proceed in the same way for all subscribers and aggregate across them. This indicator mimics the principle behind the other focal remuneration method for music streaming: namely, the \emph{user-centric rule}. 

\textbf{Subscriber-proportional indicator}. For each $(N,S,p,C) \in \mathcal{D}$ and each $i \in N$,
$$
R^{SP}_i(N,S,p,C) = \sum_{s \in S} \dfrac{C_{is}}{\|C_{\cdot s}\|} p_s.
$$

In the following example, the relevance of each service for each of the indicators defined in this section is illustrated.

\begin{example}
  Consider the platform where $N=\{1,2,3\}$, $S=\{1,2,3,4,5,6\}$, $p=\left(2,4,\frac{5}{2},2,1,\frac{7}{2} \right)$, and $C$ is given by
  $$
  \left(
  \begin{array}{cccccc}
    0 & 5 & 0 & 1 & 2 & 3 \\
    1 & 1 & 2 & 3 & 6 & 0 \\
    0 & 0 & 0 & 0 & 0 & 0 \\
  \end{array}
  \right)
  $$
  Notice that, in this case, the success is $\sigma=15$.The next table shows the relevance of each service according to the three previous indicators:
\begin{center}
\begin{tabular}{lccc}
\toprule
& \multicolumn{3}{c}{Services} \\ \cmidrule{2-4}
Indicator & 1 & 2 & 3 \\
\midrule
$R^U$ & 5 & 5 & 5 \\
$R^{SU}$ & 7 & 8 & 0 \\
$R^P$ & $\frac{105}{16}$ & $\frac{135}{16}$ & 0 \\
$R^{SP}$ & $\frac{13}{2}$ & $\frac{17}{2}$ & 0 \\
\bottomrule
\end{tabular}
\end{center}
\end{example}

\section{Normative framework}\label{axioms}

The axiomatic approach to analyzing allocation mechanisms is one of the common approaches for determining what is most appropriate, depending on the properties that are considered most relevant to the problem or context in question. In this paper, we consider some basic axioms of the literature taken from the normative framework provided in \cite{Ju07} for a wide range of resource allocation problems. Some of these axioms have been conveniently adapted to the context of the problem studied in this paper, while others are new. The first axiom requires that the sum of the success of all services is equal to the total success of the platform. We impose this axiom in the very definition of indicator, as follows:

\textbf{Efficiency}. For each $(N,S,p,C)$, $\sum_{i\in N}R_i(N,S,p,C)=\sigma$.

The following two axioms are minimum requirements of impartiality when considering different levels of information. They both require symmetric services to be treated symmetrically, but they differ on when two services should be considered as such. Thus, the first property states that two services are symmetric if each subscriber consumes one of them, as long as she does so with the other service, regardless of the consumption intensity. \emph{Strong symmetry} requires that two symmetric services have the same relevance to the success of the platform.

\textbf{Strong Symmetry}. For each $(N,S,p,C)$ and each pair $\{i,j\} \subseteq N$, if $C_{is} >0 \Leftrightarrow C_{js} >0$ for all $s\in S$, then   $R_i(N,S,p,C)=R_j(N,S,p,C)$.

As an alternative to the previous formulation, the next axiom considers that two services are symmetric if they are equally consumed by all subscribers. In such a case, both services must be equally relevant. 

\textbf{Symmetry}. For each $(N,S,p,C)$ and each pair $\{i,j\} \subseteq N$, if $C_{is}=C_{js}$ for all $s \in S$, then   $R_i(N,S,p,C)=R_j(N,S,p,C)$.

Obviously, \emph{strong symmetry} implies \emph{symmetry}. The distinction between these two properties is particularly relevant in this setting, since some indicators may satisfy the milder axiom of not the stronger one. This is a peculiarity of our model. In alternative contexts (for example, in museum pass problems (\cite{Ginsburgh03}, \cite{Martinez23}) or in risk factor problems in epidemiology (\cite{Martinez21}, \cite{Martinez22d}), among others.) \emph{strong symmetry} becomes the same as the property of \emph{symmetry}.

The following axiom states that if a service is not consumed by any subscriber, then it has no relevance within the platform, which means that its relevance is equal to zero:

\textbf{Nullity}. For each $(N,S,p,C) \in \mathcal{D}$, and each $i\in N$ such that $C_{is}=0$ for all $s\in S$, then $R_i(N,S,p,C)=0$.

The next axiom states that changing units or magnitudes does not impact the relevance of services in determining their success on the platform. In simpler terms, multiplying the consumption matrix by a positive number does not affect the relevance of the services.

\textbf{Homogeneity}. For each $(N,S,p,C)$ and each $\lambda \in \mathbb{R}_{++}$, $R(N,S,p,\lambda C)=R(N,S,p,C)$.

Suppose that there are two disjoint groups of subscribers $S$ and $S'$ (e.g., subscribers based in two regions of a country). Now consider a larger society resulting from combining $S$ and $S'$. The question is how to recalculate the indicator for $S \cup S'$ from the indicators for $S$ and $S'$. \emph{Composition} states that the relevance to the success of each service in the large population, $S \cup S'$, is the sum of the relevance in $S$ and $S'$.\footnote{This property is related to the additivity or decomposability into parts of a problem, which is also common in the resource allocation literature (see, for example, \cite{Ju07, Bergantinos15, Algaba19b, Martinez22d, Lopez23, Martinez21, Martinez23}).}

\textbf{Composition}. For each $(N,S,p,C), (N,S',p',C') \in \mathcal{D}$ such that $S \cap S' = \emptyset$,
$$
R(N,S \cup S',p \oplus p' ,C \oplus C') = R(N,S,p,C) + R(N,S',p',C'),
$$
where $C \oplus C'$ is the matrix resulting from concatenating $C$ and $C'$ by rows, and $p \oplus p'$ is the vector resulting from concatenating $p$ and $p'$.

The upcoming axiom, which we will term \emph{consumption sensitivity} states that minor alterations in the consumption matrix of the problem do not result in significant changes in the relevance of the services.

\textbf{Consumption sensitivity}. For each $\delta \geq 0$, if $\|C -C'\| < \delta$, then $\|R(N,S,p,C) - R(N,S,p,C')\| < \varepsilon(\delta)$, so that when $\delta$ goes to $0$, then $\varepsilon(\delta)$ also goes to $0$.

This property implies that if we have two successions of problems $\{(N,S,p,C^n)\}$ and $\{(N,S,p,C'^n)\}$, such that for every $\delta >0$, there exists $n_0$ such that $\forall n \geq n_0$, $\|C^n - C'^n\|<\delta$, then $\|R(N,S,p,C^n) - R(N,S,p,C'^n)\| < \varepsilon(\delta)$. Therefore, as $\|C^n - C'^n\|$ goes to $0$, $\|R(N,S,p,C^n) - R(N,S,p,C'^n)\|$ also goes to $0$. Notice that consumption sensitivity differs from the \emph{continuity} requirement that is used in the resource allocation literature. In fact, the proportional indicator, for example, satisfies continuity but violates consumption sensitivity (see Appendix A). 

Furthermore, this property should not be confused with the continuity property that is usually used in the resource allocation literature. The continuity property states that if there is a sequence of problems convergent to a given one, then the associated sequence of allocations converges to the allocation of the limit problem.

It is common for subscribers to streaming platforms to share their subscriptions among friends and relatives. They split the subscription price so that each one can access the platforms services. \textit{Sharing proofness} requires the indicator not to be altered by this type of behavior.

\textbf{Sharing proofness}. For each $(N,S,p,C) \in \mathcal{D}$, each non-empty $S' \subseteq S$, and each $s \in S'$, where $p'_s=\sum_{t \in S'}p_t$ and $C'_{is}=\sum_{t \in S'} C_{it}$ for all $i \in N$, then
$$
R(N,S,p,C)=R\left(N,\{s\} \cup S \backslash S',(p'_s,p_{S \backslash S'}),(C'_{\cdot s},C_{S \backslash S'}) \right),
$$
where $(C'_{\cdot s},C_{S \backslash S'})$ is the matrix whose entries are equal to $(C'_{\cdot s},C_{S \backslash S'})_{is}=C'_{is}$ and $(C'_{\cdot s},C_{S \backslash S'})_{it}=C_{it}$ for all $t\in S\backslash S'$, all $i\in N$ and $s\in S'$.

Our next axiom states that the relevance to success cannot be manipulated by merging or splitting services. In other words, this implies that no group of services can enhance its success on the platform by consolidating their consumption, and similarly no individual service can increase its success by creating additional services and dividing its consumption among them. This axiom was introduced by \cite{Oneill82} in the context of bankruptcy problems and recurrently used in resource allocation problems. For instance, see \cite{Ju07}.

\textbf{Non-manipulability}. For each $(N,S,p,C) \in \mathcal{D}$, each non-empty $N' \subseteq N$, and each $i \in N'$, where $C'_{is}=\sum_{j \in N'} C_{js}$ for all $s \in S$, then
$$
R_i \left(\{i\} \cup N \backslash N',S,p,(C'_{i \cdot},C_{N \backslash N'}) \right) =\sum_{j \in N'} R_j(N,S,p,C),
$$
where $(C'_{i \cdot},C_{N \backslash N'})$ is the matrix whose entries are equal to $(C'_{i \cdot},C_{N \backslash N'})_{is}=C'_{is}$, and $(C'_{i \cdot},C_{N \backslash N'})_{js}=C_{js}$ for all $s\in S$, all $j\in N\backslash N'$, and $i\in N'$.

Another way to analyze allocation mechanisms is from the perspective of game theory. To do so, associated with each problem, a game is defined that describes in some way what you want to distribute, in this case, the relevance to platform success. Once we have the game, we analyze whether the allocation mechanism coincides with any known solution from game theory, usually the Shapley value, as this is perhaps one of the most relevant (\cite{Shapley53}, \cite{Roth88}, \cite{Algaba19}). For the same problem, there may be different ways of defining a cooperative game associated with it, depending on the approach or idea you have about how a coalition generates value, in our case relevance. Next, various cooperative games are introduced and an analysis is performed to identify which of the proposed indicators -- if any -- the Shapley value of these games coincides with. Before analyzing the possible associated cooperative games, the mathematical expression of the Shapley value (\cite{Shapley53}) is presented below for completeness. A cooperative game is defined by a pair $(N,v)$, where $N$ is a finite set of players and $v$ is a function from the set of all possible subsets of $N$ to $\mathbb{R}$, such that $v(\varnothing)=0$, and the Shapley value of the game is given by

$$
Sh_i(N,v) = \sum_{S \subset N \backslash \{i\}}\frac{|S|\cdot (|N|-|S|-1)}{|N|!}\left(v(S\cup \{i\}) - v(S)\right),\quad i \in N.
$$

 Given a problem $(N,S,p,C) \in \mathcal{D}$, a cooperative game $(N,v)$ can be associated with it, where the set of players $N$ is equal to the set of services, and $v$ is the characteristic function that measures the relevance of coalition $S$ on the success of the platform. In all games that we introduce, the set of players remains the same; the only difference among them is the characteristic function. First, if we consider a kind of ``essentiality" property of the services, which means that if a service leaves the platform, all subscribers who view on that service will cease to subscribe, then the characteristic function of the associated cooperative game is defined as $v^{SU}(R)=\displaystyle\sum_{s\in N_R}p_s$ where $N_R=\{s\in S: C_{js} = 0 \: \forall j\in N\backslash R\}$ for all $R\subseteq N$. Following \cite{Ginsburgh03}, it is easy to check that the Shapley value of this game coincides with the \emph{subscriber-uniform indicator}.

From a different perspective, if we consider a kind of ``inessentiality" property of the services, which means that all subscribers are subscribed to the platform regardless of the services and the time they view them, then we can define the characteristic function of the associated cooperative game as $v^U(R)=\sigma$ for all $R\subseteq N$. In this new game, since the game remains constant for all coalitions, it is easy to observe that the Shapley value of this game coincides with the \emph{uniform indicator}.

The two previous games are extreme cases in terms of considering the ``essentiality" of a service in the success of the platform. However, they can also be considered intermediate cases when the subscription prices paid by consumers are taken as divisible. Thus, we consider the other three cooperative games, which take into account both the consumption matrix and the fact that the subscription prices are divisible. In the first one, we consider that the subscription prices can be divided among the services that subscribers consume, regardless of the time consumed for each service. In this case, the characteristic function $v$ is defined as $v^{SU}(R)=\displaystyle\sum_{s\in S}\frac{|supp(C_{Rs})|}{|supp(C_{\cdot s})|}p_s$ for all $R\subseteq N$, where $C_{Rs} = \left(C_{is}\right)_{i \in R}$ and for any $x\in\mathbb{R}^n$, $n\in\mathbb{N}_{++}$, $supp(x)=\{x_i: x_i>0\}$. It is easy to see that for all $R\subseteq N$ $v^{SU}(R)=\displaystyle\sum_{i\in R} v^{SU}(i)$, the Shapley value coincides with \emph{subscriber-uniform indicator}. Finally, when -- in addition -- the time consumed by the subscribers of each service is taken into account, then two other games can be defined. These two cooperative games can be seen as more ad hoc, but they still enable us to obtain our indicators using the Shapley value. The characteristic function of the first of them is defined as $v^{SP}(R)=\displaystyle\sum_{i\in R}\sum_{s\in S}\frac{C_{is}}{C_{\cdot s}}p_s$ for all $R\subseteq N$; in other words, the relative weight of the coalition $R$ in the price paid by each subscriber. It is easy to see that for all $R\subseteq N$ $v^{SP}(R)=\displaystyle\sum_{i\in R} v^{SP}(i)$, and then the Shapley value coincides with the \emph{subscriber-proportional indicator}. The characteristic function of the second and last game is defined as $v^P(R)=\displaystyle\sum_{i\in R}\frac{\|C_{i\cdot}\|}{\sum{j\in N}\|C_{j\cdot}\|}\sigma$ for all $R\subseteq N$; in other words, the relative weight of the coalition in the overall success. Once again, it is easy to see that for all $R\subseteq N$ $v^P(R)=\displaystyle\sum_{i\in R} v^P(i)$, and then the Shapley value coincides with the \emph{proportional indicator}.

Therefore, all the indicators that are introduced coincide with the Shapley value of some cooperative game whose characteristic function measures how each subset of services contributes to the success of the platform.

\section{Normative analysis}

In this section, we present our main results, which are characterizations of the four indicators presented in Section \ref{model}. These characterizations are given in terms of the axioms introduced in the previous section and, in some way, show what properties are associated with each of the indicators proposed to measure the relevance of the services to platform success. The next theorem states that the \emph{subscriber-proportional indicator} is the unique indicator that fulfills both \emph{composition} and \emph{non-manipulability}. Based on the connection between the \emph{subscriber-proportional indicator} and the \emph{user-centric principle}, Theorem \ref{thm1} also provides instrumental and behavioral justifications for the application of the latter, given that it provides insights into its calculation and the prevention of strategic behaviors to manipulate the measure of the success.

\begin{theorem}\label{thm1}
An indicator satisfies composition and non-manipulability if and only if it is the subscriber-proportional indicator.
\begin{proof}
	We start by showing that the subscriber-proportional indicator satisfies the axioms in the statement.
	\begin{itemize}
	  \item Composition. Let $(N,S,p,C), (N,S',p',C') \in \mathcal{D}$ such that $S \cap S' = \emptyset$. Let $i \in N$. We have that
   \begin{align*}
			R_i^{SP}(N,S \cup S',p \oplus p' ,C \oplus C') &= \sum_{s \in S \cup S'} \dfrac{(C \oplus C')_{is}}{\|(C \oplus C')_{\cdot s}\|} (p \oplus p')_s \\
			&= \sum_{s \in S} \dfrac{C_{is}}{\|C_{\cdot s}\|} p_s + \sum_{s \in S'} \dfrac{C'_{is}}{\|C'_{\cdot s}\|} p'_s  \\
			&= R_i^{SP}(N,S,p,C) + R_i^{SP}(N,S',p',C').
	\end{align*}
	  \item Non-manipulability. Let $(N,S,p,C) \in \mathcal{D}$, and let $N' \subseteq N$ and $i \in N'$ such that $C'_{is}=\sum_{j \in N'} C_{js}$ for all $s \in S$. Then
    \begin{align*}
    R^{SP}_i \left(\{i\} \cup N \backslash N',S,p,(C'_{i \cdot},C_{N \backslash N'}) \right) &= \sum_{s \in S} \dfrac{C'_{is}}{C'_{is} + \sum_{k \in N \backslash N'} C_{ks}} p_s \\
     &= \sum_{s \in S} \dfrac{\sum_{j \in N'} C_{js}}{\sum_{j \in N'} C_{js} + \sum_{k \in N \backslash N'} C_{ks}} p_s \\
     &= \sum_{s \in S} \dfrac{\sum_{j \in N'} C_{js}}{\sum_{j \in N} C_{js}} p_s \\
     &= \sum_{j \in N'} \sum_{s \in S} \dfrac{C_{js}}{\|C_{\cdot s\|}} p_s \\
     &= \sum_{j \in N'} R^{SP}_j(N,S,p,C). 
    \end{align*}
	\end{itemize}
	Now, we prove the converse. Let $R$ be an indicator that fulfills composition and non-manipulability. Let $(N,S,p,C) \in \mathcal{D}$. Suppose that $S$ is a singleton (i.e., there is only one subscriber $S=\{s\}$), and let $p^{(s)}$ and $C^{(s)}=\left( C_{1s}^{(s)}, \ldots, C_{|N|s}^{(s)} \right)$ denote the corresponding subscription price and consumption matrix of this platform, respectively. Since $R$ satisfies \emph{non-manipulability}, we can apply Theorem 9 from \cite{Ju07}\footnote{It is easy to check that when $|S|=1$, our model coincides with the resource allocation model proposed by \cite{Ju07}.} to obtain that, for each $i \in N$,
	$$
	R_i \left( N,\{s\},p^{(s)},C^{(s)} \right) = \frac{C^{(s)}_{is}}{\sum_{i \in N} C^{(s)}_{is}} p^{(s)} = R^{SP}_i \left( N,\{s\},p^{(s)},C^{(s)} \right).
	$$
	Now, suppose that $S$ is such that $|S| \geq 2$. Notice that
	$$
	p=p^{(1)} \oplus \ldots \oplus p^{(|S|)} \text{ and } C = C^{(1)} \oplus \ldots \oplus C^{(|S|)},
	$$
	where each pair $p^{(s)}$ and $C^{(s)}$ corresponds to the subscription price and the consumption matrix with only one subscriber. Since $R$ satisfies \emph{composition}, it follows that, for each $i \in N$,
	$$
	R_i \left( N,S,p,C \right) = \sum_{s \in S} R_i \left( N,\{s\},p^{(s)},C^{(s)} \right) = \sum_{s \in S} R^{SP}_i \left( N,\{s\},p^{(s)},C^{(s)} \right) = R^{SP}_i \left( N,S,p,C \right).
	$$
\end{proof}
\end{theorem}

As the next remark shows, Theorem \ref{thm1} is tight, and both axioms are necessary for the characterization.

\begin{remark}
The axioms of Theorem \ref{thm1} are independent.
\begin{itemize}
    \item[(a)] The uniform indicator satisfies composition, but not non-manipulability.
    \item[(b)] The proportional indicator satisfies non-manipulability, but not composition.
\end{itemize}
\end{remark}

Our second characterization shows that if we require \emph{sharing proofness} and \emph{non-manipulability} to be satisfied, then the relevance of each service must be measured using the \emph{proportional indicator}. As mentioned in the introduction, in our setting, the \emph{proportional indicator} implements the \emph{pro-rate principle}. Therefore, Theorem \ref{thm2} can also be interpreted as a way to justify, from a normative perspective, the application of this principle in the music streaming industry.

\begin{theorem}\label{thm2}
An indicator satisfies sharing proofness and non-manipulability if and only if it is the proportional indicator.
\begin{proof}
	We start by showing that the proportional indicator satisfies the axioms in the statement.
	\begin{itemize}
	  \item Sharing proofness. Let $(N,S,p,C) \in \mathcal{D}$. Let $S' \subseteq S$ and $s \in S'$ such that $p'_s=\sum_{t\in S'}p_t$ and $C'_{is}=\sum_{t \in S'} C_{it}$ for all $i\in N$ and $C'_{is}=C_{is}$ for all $i\in N$ and $s\in S\backslash S'$. Then
\begin{align*}
    R^P_i \left( N ,\{s\}\cup S\backslash S',(p'_s,p_{S\backslash S'}),C' \right) &= \dfrac{\|C'_{i \cdot}\|}{ \sum_{j \in N} \|C'_{j \cdot}\|} \sigma \\
    &=\dfrac{\sum_{t\in \{s\}\cup S\backslash S'}C'_{i t}}{ \sum_{j \in N}\sum_{t\in \{s\}\cup S\backslash S'} C'_{j t}} \sigma\\
    &=\dfrac{C'_{i s}+\sum_{t\in S\backslash S'}C'_{i t}}{ \sum_{j \in N} \left(C'_{j s}+\sum_{t\in S\backslash S'} C'_{j t}\right)} \sigma\\
    &=\dfrac{\sum_{t \in S'} C_{it}+\sum_{t\in S\backslash S'}C_{i t}}{ \sum_{j \in N} \left(\sum_{t \in S'} C_{jt}+\sum_{t\in S\backslash S'} C_{j t}\right)} \sigma\\
    &=\dfrac{\sum_{t \in S} C_{it}}{ \sum_{j \in N} \sum_{t \in S} C_{jt}} \sigma\\
    &= R^{P}_i(N,S,p,C). 
    \end{align*}

	  \item Non-manipulability.  Let $(N,S,p,C) \in \mathcal{D}$. Let $N' \subseteq N$ and $i \in N'$ such that $C'_{is}=\sum_{j \in N'} C_{js}$ for all $s \in S$. Then
    \begin{align*}
    R^P_i \left(\{i\} \cup N \backslash N',S,p,(C'_{i \cdot},C_{N \backslash N'}) \right) &= \dfrac{\|C'_{i \cdot}\|}{\|C'_{i \cdot}\| + \sum_{k \in N \backslash N'} \|C_{k \cdot}\|} \sigma \\
     &= \dfrac{\sum_{j \in N'} \|C_{j\cdot}\|}{\sum_{j \in N'} \|C_{j\cdot}\| + \sum_{k \in N \backslash N'} \|C_{k \cdot}\|} \sigma \\
     &= \sum_{j \in N'} \dfrac{\|C_{j\cdot}\|}{\sum_{k \in N} \|C_{k \cdot}\|} \sigma \\
     &= \sum_{j \in N'} R^{P}_j(N,S,p,C). 
    \end{align*}
	\end{itemize}
	Now, we prove the converse. Let $R$ be an indicator that fulfills sharing proofness and non-manipulability. Let $(N,S,p,C) \in \mathcal{D}$. Now, consider the platform $(N,\{1\},p^{(1)},C^{(1)})$ where $p^{(1)}=p_1+\ldots+p_{|S|}$ and $C^{(1)}_i=\sum_{s \in S} C_{is}$ for all $i\in N$. \emph{Sharing proofness} requires that
	$$
	R(N,S,p,C) = R \left( N,\{1\},p^{(1)},C^{(1)} \right).
	$$
	Notice that in $\left(N,\{1\},p^{(1)},C^{(1)}\right)$ there is only one subscriber. Applying Theorem 9 in \cite{Ju07}, \emph{non-manipulability} implies that for each $i \in N$,
	$$
	R_i\left(N,\{1\},p^{(1)},C^{(1)}\right) = \frac{C^{(1)}_i}{\|C^{(1)}\|}\sigma = \frac{\|C_{i\cdot}\|}{\|C\|} \sigma.
	$$
    Therefore, for each $i \in N$, we conclude that
    $$
    R_i(N,S,p,C) = \frac{\|C_{i\cdot}\|}{\|C\|} \sigma = R^P_i(N,S,p,C).
    $$
\end{proof}
\end{theorem}

\begin{remark}
The axioms of Theorem \ref{thm2} are independent.
\begin{itemize}
    \item[(a)] The uniform indicator satisfies sharing proofness, but not non-manipulability.
    \item[(b)] The subscriber-proportional indicator satisfies non-manipulability, but not sharing proofness.
\end{itemize}
\end{remark}

Our third characterization is given in Theorem \ref{thm3}. This result states that the combination of \emph{symmetry}, \emph{homogeneity}, and \emph{consumption sensitivity} unambiguously yield to the application of the \emph{uniform indicator}.

\begin{theorem}\label{thm3}
An indicator satisfies symmetry, homogeneity, and consumption sensitivity if and only if it is the uniform indicator.
\begin{proof}
 
 We start by showing that the uniform indicator satisfies the axioms in the statement.
 
 \begin{itemize}
  \item Symmetry. Let $(N,S,p,C) \in \mathcal{D}$. Let $i,j \in N$ such that $C_{is}=C_{js}$ for all $s\in S$. It follows that
   \begin{align*}
			R_i^{U}(N,S,p,C) &=\frac{\sigma}{|N|}=R_j^{U}(N,S,p,C). 
	\end{align*}
 \item Homogeneity. Let $(N,S,p,C) \in \mathcal{D}$ and $\lambda \in  \mathbb{R}_{++}$.  Let $i\in N$. It follows that
 \begin{align*}
			R_i^{U}(N,S,p,C) &=\frac{\sigma}{|N|}=R_i^{U}(N,S,p,\lambda C). 
	\end{align*}
\item Consumption sensitivity. Since $R_i^{U}(N,S,p,C)=\frac{\sigma}{|N|}$ for all $i \in N$, it is easy to see that small changes in the consumption matrix do not alter the relevance of the services given by the uniform indicator.
\end{itemize}

Next, we prove the converse. Let $R$ be an indicator that satisfies symmetry, homogeneity, and consumption sensitivity, and $R \neq R^U$. Therefore, there exists a problem $(N,S,p,C)$ such that $R(N,S,p,C) \neq R^U(N,S,p,C)$.

For an arbitrary $n \in \mathbb{N}_{++}$, let $C^{0n}$ be the consumption matrix for which all coordinates are exactly equal to $\frac{1}{n}$. On the one hand, by \emph{symmetry} we know that $R_i(N,S,p,C^{0n})=\frac{\sigma}{|N|}$ for all $i \in N$ and for all $n \in \mathbb{N}_{++}$. On the other hand, by \emph{homogeneity}, $R(N,S,p,\frac{1}{n}C)=R(N,S,p,C)$ for all $n \in \mathbb{N}_{++}$.

Now, it is straightforward to prove that $\|C^{0n} - \frac{1}{n}C\|$ goes to $0$ when $n$ goes to infinity, but since $R(N,S,p,C) \neq R^U(N,S,p,C)$, it follows that $\|R(N,S,p,C^{0n})- R(N,S,p,\frac{1}{n}C)\| =\|R^U(N,S,p,C^{0n})- R(N,S,p,\frac{1}{n}C)\|= k >>0$ for all $n \in \mathbb{N}_{++}$. This contradicts the fact that $R$ satisfies consumption sensitivity.

\end{proof}
\end{theorem}

\begin{remark}
The axioms of Theorem \ref{thm3} are independent.
\begin{itemize}
    \item[(a)] The proportional indicator satisfies symmetry and homogeneity, but not consumption sensitivity. To see this, consider a situation with three services and only one subscriber. Let us consider the following two successions of consumption matrices:
    $$
    C^{1n}=\left(
    \begin{array}{c}
     \frac{1}{n}\\
     \frac{1}{n}\\
     \frac{1}{n}\\
    \end{array}
    \right)
    \text{ and }
    C^{2n}=\left(
    \begin{array}{c}
     0\\
     \frac{1}{n}\\
     \frac{2}{n}\\
    \end{array}
    \right).
    $$
    It is obvious that $\|C^{1n} - C^{2n}\|$ goes to $0$ when $n$ goes to infinity, but $\|R^P(N,S,p,C^{1n})- R^P(N,S,p,C^{2n})\|=\|(\frac{p}{3},\frac{p}{3},\frac{p}{3}) - (0,\frac{p}{3},\frac{2p}{3})\| = \frac{2p}{3} >> 0$ for all $n$.
    \item[(b)] Let $R^1$ be defined as follows. For each $i \in N$, 
    $$
    R^1_i(N,S,p,C)=\frac{\sigma}{|N|}-\displaystyle \min \left\{\frac{1}{\|C_{i\cdot}\|},\|C_{i\cdot}\| \right\}+\frac{\displaystyle\sum_{j\in N}\min \left\{ \frac{1}{\|C_{j\cdot}\|},\|C_{j\cdot}\|\right\} }{|N|}.
    $$
    The indicator $R^1$ satisfies symmetry and consumption sensitivity, but not homogeneity.
    \item[(c)] Let $R^2$ be defined as follows. For each $i \in N$, 
    $$
    R^2_i(N,S,p,C)=
    \begin{cases}
    \dfrac{2\sigma}{|N|} & \text{if } i=1 \\[0.4cm]   
    \dfrac{(|N|-2)\sigma}{|N|(|N|-1)} & \text{otherwise.}
    \end{cases}
    $$
     The indicator $R^2$ satisfies homogeneity and consumption sensitivity, but not symmetry.
\end{itemize}
\end{remark}

The final characterization shows that if \emph{composition}, \emph{nullity}, and \emph{strong symmetry} must be satisfied for an indicator, then the relevance of each service must be measured using the \emph{subscriber-uniform indicator}. 

\begin{theorem}\label{thm4}
An indicator satisfies composition, nullity, and strong symmetry if and only if it is the subscriber-uniform indicator.
\begin{proof}
	We start by showing that the \emph{subscriber-proportional indicator} satisfies the axioms in the statement.
	\begin{itemize}
	  \item Composition. Let $(N,S,p,C), (N,S',p',C') \in \mathcal{D}$ such that $S \cap S' = \emptyset$. Let $i \in N$. It follows that
   \begin{align*}
			R_i^{SU}(N,S \cup S',p \oplus p' ,C \oplus C') &= \sum_{s \in S \cup S', i\in N_s}  \dfrac{1}{\left| N_s\right|} (p \oplus p')_s \\
			&= \sum_{s \in S, i\in N_s}  \dfrac{1}{\left| N_s\right|} p_s + \sum_{s \in S', i\in N_s}  \dfrac{1}{\left| N_s\right|} p'_s  \\
			&= R_i^{SU}(N,S,p,C) + R_i^{SP}(N,S',p',C').
	\end{align*}
        \item Nullity. Let $(N,S,p,C) \in \mathcal{D}$ and $i\in N$ such that $C_{is}=0$ for all $s\in S$, then $i\notin N_s=\{j\in N: C_{js}\neq 0\}$ for all $s\in S$ and 
        \begin{align*}
			R_i^{SU}(N,S ,p  ,C) &= \sum_{s \in S, i\in N_s }  \dfrac{1}{\left| N_s\right|} p_s=0.
	\end{align*}
        \item Strong symmetry. Let $(N,S,p,C) \in \mathcal{D}$. Let $i,j \in N$ such that $C_{is}\cdot C_{js}\neq 0$ or $C_{is}=C_{js}=0$ for all $s\in S$. Then $i\in N_s$ if and only if $j\in N_s$ for all $s\in S$, and it follows that
   \begin{align*}
			R_i^{SU}(N,S ,p  ,C) &= \sum_{s \in S, i\in N_s }  \dfrac{1}{\left| N_s\right|} p_s=\sum_{s \in S, j\in N_s }  \dfrac{1}{\left| N_s\right|} p_s=R_j^{SU}(N,S ,p  ,C). 
	\end{align*}
\end{itemize}

	Now, we prove the converse. Let $R$ be an indicator that fulfills \emph{composition}, \emph{nullity}, and \emph{strong symmetry}. Let $(N,S,p,C) \in \mathcal{D}$. Suppose that $S$ is a singleton (i.e., there is only one subscriber $S=\{s\}$), and let $p^{(s)}$ and $C^{(s)}=\left( C_{1s}^{(s)}, \ldots, C_{|N|s}^{(s)} \right)$ be the corresponding subscription price and consumption matrix of this platform, respectively. Since $R$ satisfies \emph{nullity}, it follows that
    $$R_i \left( N,\{s\},p^{(s)},C^{(s)} \right)=0,$$
    for all $i\in N$ such that $C^{(s)}_{is}=0$. Additionally, since $R$ satisfies \emph{strong symmetry}, it follows that
    $$R_i \left( N,\{s\},p^{(s)},C^{(s)} \right)=\frac{p^{(s)}}{|N_s|},$$
    for all $i\in N$ such that $C^{(s)}_{is}>0$, where $N_{s}=\{j\in N: C^{(s)}_{js}>0\}$.
    
	Now, suppose that $S$ is such that $|S| \geq 2$. Notice that
	$$
	p=p^{(1)} \oplus \ldots \oplus p^{(|S|)} \text{ and } C = C^{(1)} \oplus \ldots \oplus C^{(|S|)},
	$$
	where each pair $p^{(s)}$ and $C^{(s)}$ corresponds to the subscription price and the consumption matrix with only one subscriber. Since $R$ satisfies \emph{composition}, it follows that for each $i \in N$,
	$$
	R_i \left( N,S,p,C \right) = \sum_{s \in S} R_i \left( N,\{s\},p^{(s)},C^{(s)} \right) = \sum_{s \in S} R^{SU}_i \left( N,\{s\},p^{(s)},C^{(s)} \right) = R^{SU}_i \left( N,S,p,C \right).
	$$
\end{proof}
\end{theorem}

\begin{remark}
The axioms of Theorem \ref{thm4} are independent.
\begin{itemize}
    \item[(a)] The subscriber-proportional indicator satisfies composition and nullity, but not strong symmetry. To see this, consider a situation with three services, only one subscriber, $p_1=1$, and the following consumption matrix:
      $$
    C^{1n}=\left(
    \begin{array}{c}
     2\\
    1\\
    0\\
    \end{array}
    \right)
    $$
For the strong symmetry axiom, the success of service one and two should be equal. However, $R^{SP}_1(N,S,p,C)=\frac{2}{3}\neq \frac{1}{3}=R^{SP}_2(N,S,p,C)$.
\item[(b)] The uniform indicator satisfies composition and strong symmetry, but not nullity. Consider the example of the previous item. Service three is not viewed by the subscriber, but $R^{SP}_3(N,S,p,C)=\frac{1}{3}$.
 \item[(c)] Let $R^1$ be defined as follows. For each $i \in N$, 
    $$
    R^1_i(N,S,p,C)=
    \begin{cases}
    \dfrac{\sigma}{|M|} & \text{if } M\neq \emptyset \\[0.4cm]   
    R^{P}_i(N,S,p,C) & \text{otherwise,}
    \end{cases}
    $$
    where $M\subseteq N$ such that if $i\in M$, there exists at least one service $j\in M$, $C_{is}\cdot C_{js}\neq 0$ or $C_{is}=C_{js}=0$ for all $s\in S$. 
    
    The indicator $R^1$ satisfies nullity and strong symmetry, but not composition.
\end{itemize}
\end{remark}

Table \ref{Table1} summarizes our theoretical findings. As we can observe, the four indicators satisfy several interesting properties, even though not all properties are required for their characterizations. The \emph{uniform indicator} fulfills all the axioms except \emph{non-manipulability} and \emph{nullity}. The \emph{subscriber-uniform indicator} satisfies all the axioms except \emph{consumption sensitivity}, \emph{sharing proofness}, and \emph{non-manipulabity}, and it is characterized by \emph{composition}, \emph{strong symmetry}, and \emph{nullity}. \emph{Sharing proofness} and \emph{non manipulability} unambiguously yield to the \emph{proportional indicator}, but they also satisfy \emph{symmetry}, \emph{homogeneity}, and \emph{nullity}. The \emph{subscriber-proportional indicator} is characterized by \emph{composition} and \emph{non-manipulability}, but it also fulfills \emph{symmetry}, \emph{homogeneity}, and \emph{nullity}. From the standpoint of the axiom, three axioms -- (\emph{efficiency}, \emph{symmetry}, and \emph{homogeneity}) -- are satisfied by all indicators in the paper. Moreover, all axioms except \emph{consumption sensitivity} are satisfied by at least two of the four indicators.

\begin{table}[H]
\begin{center}
\begin{tabular}{lcccc}
\toprule
Property & $R^U$& $R^{SU}$ & $R^P$ & $R^{SP}$ \\ 
\midrule
Efficiency              & Y & Y & Y & Y \\
Symmetry                & Y$^*$ & Y & Y & Y \\
Strong symmetry            & Y & Y$^*$ & N & N \\
Homogeneity             & Y$^*$ & Y & Y & Y \\
Consumption sensitivity & Y$^*$ & N & N & N \\
Composition             & Y & Y$^*$ & N & Y$^*$ \\
Sharing proofness       & Y & N & Y$^*$ & N \\
Non-manipulability      & N & N & Y$^*$ & Y$^*$ \\
Nullity                 & N & Y$^*$ & Y & Y \\
\bottomrule
\end{tabular}
\caption{Summary of theoretical findings. Y means that the indicator in the column does satisfy the axiom in the row, N means that the indicator does not satisfy the axiom, and Y$^*$ means that the axiom is used to characterize the indicator.\label{Table1}}
\end{center}
\end{table}

For proofs and counterexamples of all properties for all studied relevance indicators that have not been proven in Theorems \ref{thm1} to \ref{thm4}, see Appendix A.

\section{Analysis of the indicators in the streaming context}

In this section, we analyze relevance indicators in the context of streaming platforms. To do this, different aspects related to the information elements of the problem, the structure of the indicators, and the properties that they satisfy are considered. Likewise, we analyze which indicators are most attractive to the interests of the platform.

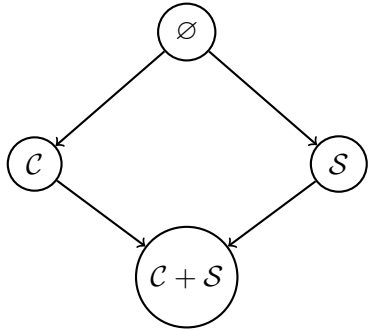
\begin{figure}[H]
\begin{center}
	\begin{tikzpicture}[node distance={25mm}, thick, main/.style = {draw, circle}] 
		\node[main] (1) at (6,0) {$\varnothing$}; 
         \node[main] (2) at (4,-1.75) {$\mathcal{C}$}; 
          \node[main] (3) at (8,-1.75) {$\mathcal{S}$}; 
          \node[main] (4) at (6,-3.25) {$\mathcal{C+S}$}; 
          \draw[->] (1) -- (2); 
          \draw[->] (1) -- (3); 
          \draw[->] (2) -- (4); 
          \draw[->] (3) -- (4); 
	\end{tikzpicture} 
\end{center}
\caption{Levels of information on a platform.} 
\label{fig.1.}
\end{figure}

Given a platform described by $\left(N,S,p,C\right)$, it is possible to distinguish four levels of the use of the information contained in the problem description. The simplest level, which we denote by $\varnothing$, would correspond to using only the number of streaming creators (streaming channels or simply streamers) on the platform as information. In this case, both the number of subscribers and their streamer consumption times are considered irrelevant and, therefore, the only important consideration is how many streamers are offered on the platform and how much is earned from subscriptions. At the second level, we consider two possibilities. On the one hand, we consider the streamers, the total time they are watched for, and the total subscription revenue as relevant information. In this case, the number of subscribers and the time that each of them watches the streamers is not taken into account. Therefore, the only relevant consideration is how much time each streamer is watched and how much is earned. We denote this level of information by $\mathcal{C}$. On the other hand, consider subscribers, the streamers they watch, and their subscription fees as relevant information. In this case, the consumption times of each subscriber of each of the streamers are not taken into account. Therefore, the only important consideration is how many subscribers watch each streamer and how much they pay. We denote this level of information by $\mathcal{S}$. The richest level in terms of information, which we denote by $\mathcal{C+S}$, would correspond to using all the information related to subscribers, the streamers they watch, their watching times, and their subscription fees. Therefore, in this case, all the information available in the problem description is used. All these levels of information are depicted in Figure \ref{fig.1.}.

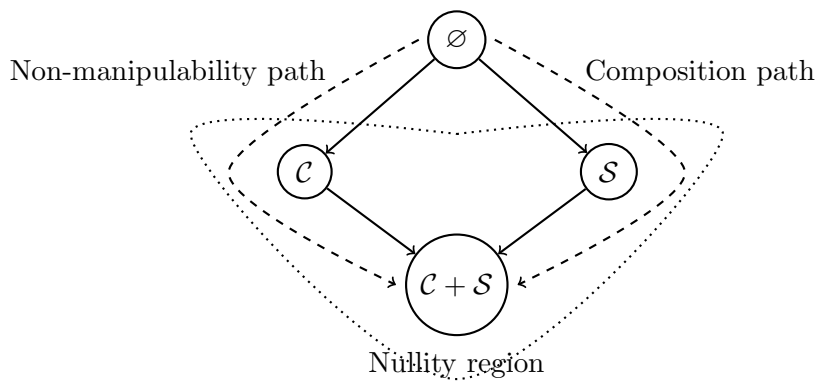
\begin{figure}[H]
\begin{center}
	\begin{tikzpicture}[node distance={25mm}, thick, main/.style = {draw, circle}] 
		\node[main] (1) at (6,0) {$\varnothing$}; 
         \node[main] (2) at (4,-1.75) {$\mathcal{C}$}; 
          \node[main] (3) at (8,-1.75) {$\mathcal{S}$}; 
          \node[main] (4) at (6,-3.25) {$\mathcal{C+S}$};
          \draw[->] (1) -- (2); 
          \draw[->] (1) -- (3); 
          \draw[->] (2) -- (4); 
          \draw[->] (3) -- (4);
          \draw[->, dashed] plot[smooth] coordinates{(6.5,0) (9,-1.75) (6.8,-3.25)};
          \draw (9.2,-0.75) node[above] {Composition path};
          \draw[->, dashed] plot[smooth] coordinates{(5.5,0) (3,-1.75) (5.2,-3.25)};
          \draw (2.2,-0.75) node[above] {Non-manipulability path};
          \draw[dotted] plot[smooth] coordinates{(6,-1.25) (9.5,-1.25) (6,-4.5) (2.5,-1.25) (6,-1.25)};
          \draw (6,-4) node[below] {Nullity region};
    \end{tikzpicture} 
\end{center}
\caption{Levels of information and properties.} 
\label{fig.2.}
\end{figure}

Below we relate each of the levels of information described in Figure \ref{fig.1.} with the relevance indicators presented in Section \ref{model} and the properties introduced in Section \ref{axioms}. Of the nine properties presented in the normative framework, two of them -- \emph{efficiency} and \emph{homogeneity} -- are present at all levels of information; they do not provide differentiating elements in the study of the relationship between the properties and the levels of information. These two properties are related to the earnings obtained by the platform through the payment of subscription fees, which are a common element in all levels of information considered. On the other hand, the properties of \emph{composition} and \emph{non-manipulability} are associated with two different paths of information aggregation. The path of \emph{non-manipulability} takes us to the information levels $\mathcal{C}$ and $\mathcal{C+S}$, which have in common that the time watching the streamers is a fundamental piece of information. In this sense, the fact of considering the watching times of the streamers would be related to the non-possibility of manipulating the relevance measures by the streamers themselves. The \emph{composition} path leads us to the information levels $\mathcal{S}$ and $\mathcal{C+S}$ that have in common that what individual subscribers watch on the platform is an essential piece of information. In this case, considering what each subscriber sees on the platform is related to the possibility of decomposing the problem of measuring relevance into smaller parts. The \emph{nullity} is related to all levels of information that take into account, in one way or another, subscribers’ behavior on the platform; therefore, this property is not associated with the information level $\varnothing$, but with the other three levels of information. Figure \ref{fig.2.} depicts the properties and levels of information discussed here.

The relationship of the four remaining properties with the different levels of information is as follows. The property of \emph{consumption sensitivity} implies that in the environment of null values, the information regarding which subscribers and how long they watch the streamers becomes almost irrelevant, as long as the subscription rates that they pay remain the same. Therefore, this property would be associated with the information level $\varnothing$. In this way, this property becomes the antagonist of \emph{nullity}, which refers to null viewing time values; however, knowing which subscribers they are and which streamers are not watched, then which subscribers and how long they watch the streamers, is relevant. The symmetry property is associated with the information level $\mathcal{C+S}$ since its very definition requires information about which streamers subscribers watch and for how long. However, the \emph{strong symmetry} property is only related to the streamers that subscribers watch, but without going into the detail of how long they watch them; therefore, this property is associated with the information level $\mathcal{S}$. It is easy to check that \emph{symmetry} implies \emph{strong symmetry}, but the converse is not true. The last property proposed in the normative framework is \emph{sharing proofness}. In this property, the only relevant element is the time that each streamer is watched, regardless of who and how many subscribers watch them, then it is related to the information level $\mathcal{C}$.  

As has been done with properties, it is also possible to relate relevance indicators to information levels. First, since the \emph{uniform indicator} exclusively relies on the number of services to distribute the platform's success, this indicator would be related to the level of information $\varnothing$. Second, the \emph{proportional indicator} only considers the total viewing time of each streamer to measure their relevance to platform success; in this way, this indicator would be related to the level of information $\mathcal{C}$. Third, the \emph{subscriber-uniform indicator} uses information regarding which subscribers watch the streamers, but not for how long they watch them, meaning that this indicator would be related to the level of information $\mathcal{S}$. Finally, in the definition of the \emph{subscriber-proportional indicator}, both the subscribers who watch the streamers and the time they watch them are taken into consideration; therefore, this indicator would be related to the level of information $\mathcal{C+S}$. In this way, through the levels of information, the indicators are also related to the different properties. At this point, it is worth saying that there may be indicators that satisfy properties related to other levels of information other than the one that the indicator is related to, as is the case in this context (see Table \ref{Table1}). The reason for this circumstance is largely due to the structure of the indicators themselves. For example, all of the presented indicators satisfy the \emph{symmetry} property. In the case of the \emph{uniform indicator}, it satisfies the \emph{symmetry} property by the simple fact that all streamers are treated identically regardless of other elements. The \emph{proportional indicator} satisfies the \emph{symmetry} property as a consequence of the fact that if all the times of two symmetric streamers are added together, their sum will obviously be equal and the indicator will treat them the same by its own definition. Finally, the \emph{subscriber-uniform indicator} satisfies the \emph{symmetry} property for the simple fact that if two streamers are symmetrical, then they will be watched by the same subscribers.

The following example illustrates when a streamer can prefer one indicator over another. This is a toy example with two subscribers, in which three types of streamers are considered: one watched by one subscriber and not watched by the other, another watched equally by both subscribers, and another not watched by either subscriber. In particular, let $(N,S,p,C) \in \mathcal{D}$, where $|N|=3$ and $|S|=2$; then the subscription prices $p=(p_1,p_2)$ and the consumption matrix is equal to
 $$
\left(
\begin{array}{cc}
	M & 0  \\
	1 & 1\\
	0 & 0 \\
\end{array}
\right),
$$
where $M\in\mathbb{R}_{++}$.

In this scenario, we have the following three types of streamers: Streamer 1 is viewed exclusively by the first subscriber, with a variable consumption quantity; Streamer 2 is viewed by the first subscriber for a single unit of time (we can normalize this unit for homogeneity with respect to what this subscriber watches of the first service), and it is exclusively viewed by the second subscriber; and Streamer 3, on the other hand, remains entirely unwatched by any subscriber. The following table shows the relevance of each streamer according to the four indicators:
\begin{center}
\begin{tabular}{lccc}
\toprule
& \multicolumn{3}{c}{Streamers} \\ \cmidrule{2-4}
Indicator & 1 & 2 & 3 \\
\midrule
$R^U$ & $\frac{(p_1+p_2)}{3}$ & $\frac{(p_1+p_2)}{3}$ & $\frac{(p_1+p_2)}{3}$ \\
$R^{SU}$ & $\frac{p_1}{2}$ & $\frac{p_1}{2} + p_2$ & 0 \\
$R^P$ & $\frac{M}{M+2}(p_1+p_2)$ & $\frac{2}{M+2}(p_1+p_2)$ & 0 \\
$R^{SP}$ & $\frac{M}{M+1}p_1$ & $\frac{1}{M+1}p_1+p_2$ & 0 \\
\bottomrule
\end{tabular}
\end{center}

It is obvious that a streamer that is not viewed (or hardly watched) by any subscriber will prefer the \emph{uniform indicator}. Since the rest of the indicators satisfy the axiom of \emph{nullity}, it would be assigned null relevance in any other case. For the other two streamers, we have the following cases :
\begin{itemize}
    \item $p_1>2p_2$ then $R^{SU}_1>R^{U}_1$.
    \item $p_1<2p_2$ then $R^{SU}_1<R^{U}_1$.
    \item For all values $R^{SU}_2>R^{U}_2$.
\end{itemize}

Since Streamer 1 is viewed exclusively by the first subscriber, it would prefer to measure its relevance using the \emph{subscriber-uniform indicator} rather than the \emph{uniform indicator}, when the first subscriber's subscription is more than twice as large as the second subscriber's. Streamer 2 will always prefer the \emph{subscriber-uniform indicator} over the \emph{uniform indicator} for any $p_1$ and $p_2$.

On the other hand, we can also analyze when a streamer prefers the \emph{proportional indicator} over the \emph{uniform indicator}:

\begin{itemize}
    \item $M>1$ then $R^{P}_1>R^{U}_1$.
    \item $M<1$ then $R^{P}_1<R^{U}_1$.
\end{itemize}

Note that Streamer 1 prefers the \emph{proportional indicator} over the \emph{uniform indicator} when its viewing time is greater than the viewing time of Streamer 2. 

\begin{itemize}
    \item $M<4$ then $R^{P}_2>R^{U}_2$.
    \item $M>4$ then $R^{P}_2<R^{U}_2$.
\end{itemize}

However, Streamer 2 will prefer the \emph{proportional indicator} to measure its relevance when the viewing time of Streamer 1 is not excessively greater (in this example, four times) than the viewing time of Streamer 2. Otherwise, Streamer 2 will prefer the \emph{uniform indicator}, since if it is much greater, Streamer 2 would be very close to the situation of Streamer 3.

We now explore when the \emph{subscriber-proportional indicator} is preferred to the \emph{uniform indicator} by Streamers 1 and 2.

\begin{itemize}
    \item $p_1\leq \frac{p_2}{2}$ then $R^{SP}_1<R^{U}_1$.
    \item $p_1>\frac{p_2}{2}$ 
    \begin{itemize}
        \item $M<\frac{p_1+p_2}{2p_1-p_2}$ then $R^{SP}_1<R^{U}_1$.
        \item $M>\frac{p_1+p_2}{2p_1-p_2}$ then $R^{SP}_1>R^{U}_1$.
    \end{itemize}
    
\end{itemize}

If the first subscriber’s subscription is half or less than that of the second subscriber, Streamer 1 prefers the \emph{uniform indicator} over the \emph{subscriber-proportional indicator}. This is because, with the \emph{subscriber-proportional indicator}, Streamer 1 does not benefit from this subscription since the second subscriber does not view Service 1. Therefore, the distribution of subscription values plays a crucial role in deciding between the indicators. If the first subscriber’s subscription is greater than that of the second, the choice of indicator will depend on the viewing time that this streamer receives; for a sufficiently high value, the chosen indicator will be the \emph{subscriber-proportional indicator}.

\begin{itemize}
    \item $p_1\leq 2p_2$ then $R^{SP}_2>R^{U}_1$.
    \item $p_1>2p_2$
    \begin{itemize} 
    \item $M<\frac{2(p_1+p_2)}{p_1-2p_2}$ then $R^{SP}_2>R^{U}_2$.
    \item $M>\frac{2(p_1+p_2)}{p_1-2p_2}$ then $R^{SP}_2<R^{U}_2$.
    \end{itemize}
\end{itemize}

If either the first subscriber’s subscription price is less than double that of the second, or the viewing time for Streamer 1 is below the threshold $\frac{2(p_1+p_2)}{p_1-2p_2}$, then Streamer 2 will opt for the \emph{subscriber-proportional indicator} over the \emph{uniform} one. This preference is due to the fact that Streamer 2 receives the entire subscription $p_2$, while also receiving a proportional share of $p_1$.

Next, we analyze when the \emph{subscriber-uniform indicator} is preferred to the \emph{proportional indicator} by Streamers 1 and 2.

\begin{itemize}
    \item $M>\frac{2p_1}{p_1+2p_2}$ then $R^P_1>R^{SU}_1$ and  $R^P_2<R^{SU}_2$.
    \item $M<\frac{2p_1}{p_1+2p_2}$ then $R^P_1<R^{SU}_1$ and  $R^P_2>R^{SU}_2$.
\end{itemize}

When the viewing time for Streamer 1 exceeds a certain threshold, denoted as $\frac{2p_1}{p_1+2p_2}$, Streamer 1 will opt for the \emph{proportional indicator}, while Streamer 2 will choose the \emph{subscriber-uniform indicator}. This decision is driven by the fact that the \emph{proportional indicator} considers the total viewing time, whereas the \emph{subscriber-uniform indicator} accounts for the number of streamers consumed by each subscriber. In this case, the second subscriber exclusively consumes Streamer 2.

We now study when the \emph{subscriber-proportional indicator} is preferred to the \emph{subscriber-uniform indicator} by Streamers 1 and 2.

\begin{itemize}
    \item $M>1$ then $R^{SP}_1>R^{SU}_1$ and  $R^{SP}_2<R^{SU}_2$.
    \item $M< 1$ then $R^{SP}_1<R^{SU}_1$ and  $R^{SP}_2>R^{SU}_2$.
\end{itemize}

When the viewing time for Streamer 1 exceeds the viewing time for Streamer 2, Streamer 1 prefers the \emph{subscriber-proportional indicator}, while Streamer 2 opts for the \emph{subscriber-uniform indicator}. This preference arises because both streamers compete for the first subscriber's subscription. With a higher viewing time, Streamer 1 receives a larger share of this subscription. As a result, Streamer 2 prefers an equal distribution, which is achieved through the \emph{subscriber-uniform indicator}.

Finally, we study when the \emph{proportional indicator} is preferred to the \emph{subscriber-proportional indicator} by Streamers 1 and 2.

\begin{itemize}
    \item $p_1<p_2$ then $R^P_1>R^{SP}_1$ and $R^P_2<R^{SP}_2$.
    \item $p_1>p_2$
    \begin{itemize}
        \item $0<M<\frac{p_1-p_2}{p_2}$ then $R^P_1<R^{SP}_1$ and $R^P_2>R^{SP}_2$.
        \item  $M>\frac{p_1-p_2}{p_2}$ then $R^P_1>R^{SP}_1$ and $R^P_2<R^{SP}_2$.
    \end{itemize}
\end{itemize}

If the first subscriber's subscription price is lower than that of the second subscriber, Streamer 1 will prefer the \emph{proportional indicator} over the \emph{subscriber-proportional indicator}. This preference is driven by the fact that the \emph{proportional indicator} allows Streamer 1 to receive a share of the second subscriber's subscription fee, whereas the \emph{subscriber-proportional indicator} does not. Conversely, Streamer 2 will opt for the \emph{subscriber-proportional indicator}, as it will receive the entirety of the second subscriber's subscription fee. If the first subscriber’s subscription price is higher, the choice of indicator will depend on the viewing time of Streamer 1.

The following lemmas are formulated for general problems that establish conditions under which a streamer will prefer one indicator over another. Proofs of all lemmas are available in Appendix B.

\begin{lemma}
\label{le.0.}
    Let $(N,S,p,C) \in \mathcal{D}$. For a streamer $i\in N$ such that $$\displaystyle\sum_{s\in S, i\in N_s}\frac{\left(|N|-|N_s|\right)}{|N_s|}p_s\geq \sum_{s\in S, i\notin N_s}p_s$$ then $R_i^{SU}(N,S,p,C)\geq R_i^{U}(N,S,p,C)$.
\end{lemma}

\begin{lemma}
\label{le.1.}
    Let $(N,S,p,C) \in \mathcal{D}$. For a streamer $i\in N$ such that $C_{is}>0$ for all $s\in S$, then $R_i^{SU}(N,S,p,C)\geq R_i^{U}(N,S,p,C)$.
\end{lemma}

Lemma \ref{le.1.} asserts that when a streamer is viewed by all subscribers, it will favor the \emph{subscriber-uniform indicator} over the \emph{uniform indicator} to measure its influence on the platform.

\begin{lemma}
\label{le.2.}
    Let $(N,S,p,C) \in \mathcal{D}$. For a streamer $i\in N$ such that $\|C_{i\cdot}\|\geq \frac{\sum_{j\in N\backslash\{i\}}\|C_{j\cdot}\|}{|N|-1}$, then $R_i^{P}(N,S,p,C)\geq R_i^{U}(N,S,p,C)$.
\end{lemma}

Lemma \ref{le.2.} states that if a streamer is viewed for more than the average viewing time of all other streamers, it will favor the \emph{proportional indicator} over the \emph{uniform indicator} to assess its influence within the platform.

\begin{lemma}
\label{le.3.}
    Let $(N,S,p,C) \in \mathcal{D}$. For a streamer $i\in N$ such that $C_{is}\geq \frac{\sum_{j\in N\backslash\{i\}}C_{js}}{|N|-1}$ for all $s\in S$ then $R_i^{SP}(N,S,p,C)\geq R_i^{U}(N,S,p,C)$.
\end{lemma}

Lemma \ref{le.3.} states that if a streamer is viewed for more than the average viewing time of all other streamers by each subscriber, then the streamer will prefer the \emph{subscriber-proportional indicator} over the \emph{uniform indicator} to measure its success on the platform.

\begin{lemma}
\label{le.4.}
    Let $(N,S,p,C) \in \mathcal{D}$. For a streamer $i\in N$ such that $C_{is}\geq \frac{\sum_{j\in N_s\backslash\{i\}}C_{js}}{|N_s|-1}$, where $ N_s=\left\{ j \in N: C_{js} \neq 0\right\}$ for all $s\in S$ with $i\in N_s$, then $R_i^{SP}(N,S,p,C)\geq R_i^{SU}(N,S,p,C)$.
\end{lemma}

Lemma \ref{le.4.} asserts that when a streamer accumulates more viewing time from each subscriber than the average viewing time on other streamers seen by the same subscriber, the streamer prefers the \emph{subscriber-proportional indicator} over the \emph{subscriber-uniform indicator}.

Let us now see how the \emph{sharing proofness} and \emph{non-manipulability} axioms affect the distribution of each indicator. To do this, let us consider the previous example again and assume that subscriber one comprises two individuals who share the subscription equally. Both have viewed Streamer 1 for the same duration, but only one of them has watched Streamer 2. Therefore, we have the following viewing time matrix:

$$
\left(
\begin{array}{ccc}
	\frac{M}{2} & \frac{M}{2} & 0\\
	1 &0 &1\\
	0 & 0&0\\
\end{array}
\right).
$$
This situation is directly related to the \emph{sharing proofness} axiom. Let us now explore how it influences the distribution of each indicator:
\begin{center}
\begin{tabular}{lccc}
\toprule
& \multicolumn{3}{c}{Streamers} \\ \cmidrule{2-4}
Indicator & 1 & 2 & 3 \\
\midrule
$R^U$ & $\frac{(p_1+p_2)}{3}$ & $\frac{(p_1+p_2)}{3}$ & $\frac{(p_1+p_2)}{3}$ \\
$R^{SU}$ & $\frac{3p_1}{4}$ & $\frac{p_1}{4}+p_2$& 0 \\
$R^P$ & $\frac{M}{M+2}(p_1+p_2)$ & $\frac{2}{M+2}(p_1+p_2)$ & 0 \\
$R^{SP}$ & $\frac{M+1}{M+2}p_1$ & $\frac{1}{M+2}p_1+p_2$ & 0 \\
\bottomrule
\end{tabular}
\end{center}

The \emph{uniform} and \emph{proportional indicators} satisfy the aforementioned axiom, so that the distribution obtained is maintained compared to the first scenario. On the other hand, the \emph{subscriber-uniform} and \emph{subscriber-proportional indicators} attributed greater relevance to Streamer 1. This is because in the first scenario, the subscription was uniformly or proportionally divided between Streamers 1 and 2 based on the viewing time. In this new scenario, half of the subscription is allocated exclusively to Streamer 1, as it is viewed solely by one of the subscribers, while the remaining half is divided either uniformly or proportionally. 

In the latter scenario, we will assume that Streamer 1 is divided into two such that the sum of their viewing times remains constant. The viewing time matrix is as follows:
$$
\left(
\begin{array}{cc}
	\frac{M}{2} & 0  \\
    \frac{M}{2} & 0  \\
	1 & 1\\
	0 & 0 \\
\end{array}
\right).
$$
This new scenario is linked to the \emph{non-manipulability} axiom. The distributions obtained with the four indicators are as follows:
\begin{center}
\begin{tabular}{lcccc}
\toprule
& \multicolumn{4}{c}{Services} \\ \cmidrule{2-5}
Indicator & $1^{1}$ & $1^{2}$ & 2 & 3 \\
\midrule
$R^U$ & $\frac{(p_1+p_2)}{4}$ & $\frac{(p_1+p_2)}{4}$ & $\frac{(p_1+p_2)}{4}$ & $\frac{(p_1+p_2)}{4}$ \\
$R^{SU}$ & $\frac{p_1}{3}$ & $\frac{p_1}{3}$ & $\frac{p_1}{3}$ + $p_2$& 0 \\
$R^P$ & $\frac{\frac{M}{2}}{M+2}(p_1+p_2)$ &$\frac{\frac{M}{2}}{M+2}(p_1+p_2)$ & $\frac{2}{M+2}(p_1+p_2)$ & 0 \\
$R^{SP}$ & $\frac{\frac{M}{2}}{M+2}p_1$ & $\frac{\frac{M}{2}}{M+2}p_1$ & $\frac{2}{M+2}p_1+p_2$ &0 \\
\bottomrule
\end{tabular}
\end{center}

As seen previously, the \emph{proportional} and \emph{subscriber-proportional indicators} satisfy the \emph{non-manipulability} axiom; therefore, their relevance remains the same as in the first scenario. On the other hand, as the number of streamers increases, both the \emph{uniform} and \emph{subscriber-uniform indicators} assign lower relevance to each streamer in the new scenario. Nevertheless, the total relevance achieved by Streamer 1 surpasses that of the initial scenario. The following lemma establishes this result for any problem:

\begin{lemma}
    \label{le.6.}
     Let $(N,S,p,C), (N',S,p,C') \in \mathcal{D}$. Let $i\in N$, and assume $N'=N\backslash\{i\}\cup\{i'_1,\hdots,i'_m\}$, with $m\geq 2$. Suppose $C_{is}=\sum_{l=1}^m C'_{i'_ls}$ and $C_{js}=C'_{js}$ for all $j\in N\backslash\{i\}$ and $s\in S$. Then, $\sum_{l=1}^mR^U_{i'_l}(N',S,p,C')\geq R^U_i(N,S,p,C)$ and $\sum_{l=1}^mR^{SU}_{i'_l}(N',S,p,C')\geq R^{SU}_i(N,S,p,C)$.
\end{lemma}

Lemma \ref{le.6.} states that if a streamer is divided into several parts by distributing its viewing time among them, it will yield greater relevance with the \emph{uniform} and \emph{subscriber-uniform indicators}.

Now, let us examine how the relevance of each streamer within the platform is affected when the viewing time of one of the streamers increases or decreases. First, since the \emph{uniform indicator} solely depends on the number of streamers, the relevance of each streamer, under this indicator, remains constant regardless of changes in the viewing time matrix. 

Second, let us see what happens with the \emph{subscriber-uniform indicator}. Let $(N,S,p,C) \in \mathcal{D}$. Given $s\in S$ and $i\in N$, suppose $C'$ such that $C'_{is}>C_{is}$ and $C'_{js'}=C_{js'}$ for all $j\in N$ and $s'\in S\backslash \{s\}$. Now, we have the following two cases:
\begin{enumerate}
    \item If $C_{is}=0$, then $R^{SU}_i(N,S,p,C')=R^{SU}_i(N,S,p,C)+\frac{p_s}{|N_s|+1}$ and $R^{SU}_j(N,S,p,C')=R^{SU}_j(N,S,p,C)-\frac{p_s}{(|N_s|+1)|N_s|}$, where $ N_s=\left\{ j \in N: C_{js} \neq 0\right\}$.
    \item If $C_{is}>0$, then $R^{SU}_i(N,S,p,C')=R^{SU}_i(N,S,p,C)$ and $R^{SU}_j(N,S,p,C')=R^{SU}_j(N,S,p,C)$, where $ N_s=\left\{ j \in N: C_{js} \neq 0\right\}$.
\end{enumerate}

Therefore, under the \emph{subscriber-uniform indicator}, streamers will not actively pursue additional views from subscribers who have already watched them, as it will not enhance their relevance on the success on the platform. Instead, the emphasis should be on subscribers who have not yet viewed the streamer, as this will yield a positive outcome. Since the allocated relevance is inversely proportional to the number of streamers viewed by a subscriber, the focus should be on those subscribers who watch the fewest services for the greatest improvements.

Third, we analyze the \emph{proportional indicator}. The mathematical formula for this indicator is given by
$$
R^P_i(N,S,p,C) = \frac{\|C_{i \cdot}\|}{\displaystyle\sum_{j \in N} \|C_{j \cdot}\|} \sigma.
$$
Then, taking the partial derivative of the previous expression with respect to $C_{is}$, we obtain that 
$$
\frac{\partial R^{P}_i(N,S,p,C)}{\partial C_{is}} = \dfrac{\displaystyle\sum_{j\in N\backslash \{i\}}\|C_{j \cdot}\|}{\displaystyle\left(\sum_{j \in N} \|C_{j \cdot}\|\right)^2} \sigma.
$$

Therefore, the improvement in the relevance of streamer $i$ when a subscriber increases the viewing time of that streamer is inversely proportional to the square of the total viewing time and directly proportional to the total viewing time of the rest of the streamers on the platform. Note that if instead of examining the viewing time of a streamer by a subscriber, we examined the total viewing time of a streamer, the result would be identical. On the other hand, if we rewrite the previous expression as
$$
\frac{\partial R^{P}_i(N,S,p,C)}{\partial C_{is}} = \left(\dfrac{1}{\displaystyle \sum_{j \in N} \|C_{j \cdot}\|}-\dfrac{\|C_{i \cdot}\|}{\displaystyle(\sum_{j \in N} \|C_{j \cdot}\|)^2}\right) \sigma,
$$
then we observe that the improvement in streamer relevance is less than the income per unit of viewing time when the viewing time of a streamer increases. Furthermore, we observe that this improvement will be less the greater the viewing time that the streamer initially had.

Fourth, we now study the changes in the \emph{subscriber-proportional indicator}. The mathematical expression for this indicator is given by
$$
R^{SP}_i(N,S,p,C) = \sum_{s \in S} \dfrac{C_{is}}{\|C_{\cdot s}\|} p_s.
$$
Then, taking the partial derivative of the previous expression with respect to $C_{is}$, it follows that
$$
\frac{\partial R^{SP}_i(N,S,p,C)}{\partial C_{is}} =  \dfrac{\displaystyle\sum_{j\in N\backslash \{i\}}C_{js}}{\|C_{\cdot s}\|^2} p_s.
$$

Therefore, we obtain a result similar to that of the \emph{proportional indicator}. In this case, the improvement in the relevance of streamer $i$ when a subscriber increases the viewing time of that streamer is inversely proportional to the square of the total viewing time of that particular subscriber and directly proportional to the total viewing time of the rest of the streamers on the platform, again of that particular subscriber. As in the case of the \emph{proportional indicator}, if we rewrite the previous expression as
$$
\frac{\partial R^{SP}_i(N,S,p,C)}{\partial C_{is}} = \left(\dfrac{1}{\displaystyle \|C_{\cdot s}\|}-\dfrac{C_{is}}{\displaystyle \|C_{\cdot s}\|^2}\right) p_s,
$$
then we observe the following: that the improvement of the relevance of a streamer is less than the subscription fee per unit of viewing time when the viewing time of that streamer by the subscriber increases; and that this improvement will be less the greater the viewing time that the streamer initially had for that subscriber. In essence, the interpretation is analogous to that obtained for the \emph{proportional indicator}, but considering only the subscription rate of the particular subscriber.

Finally, we analyze which indicator would be chosen (or preferred) by the platform. To do this, we must consider how a platform's success is measured. It can be assessed that success is determined by the number of subscriptions; the more subscribers a platform has, the higher its profits. The indicators that are directly linked to the number of subscribers are the \emph{subscriber-uniform indicator} and the \emph{subscriber-proportional indicator}. The \emph{subscriber-uniform indicator} does not consider the viewing time from each subscriber. Consequently, each streamer constantly aims to attract new subscribers, regardless of the viewing time. This leads to the addition of new subscribers to the platform, which in turn increases the platform's success and profits. On the other hand, when utilizing the \emph{subscriber-proportional indicator}, streamers seek to attract new subscribers, although to a lesser degree compared to the \emph{subscriber-uniform indicator}. Under this new metric, streamers also strive to retain these new subscribers, with the goal of prolonging their viewing time. Let us now contemplate a scenario in which the number of subscribers remains fixed, and the platform derives benefits (for instance, through advertising revenue) from the time that subscribers spend viewing streamers. In such a situation, the chosen indicators for the platform would be the \emph{proportional indicator} and the \emph{subscriber-proportional indicator}. Although the \emph{subscriber-proportional indicator} aims to achieve a balance between the number of subscribers and the time they spend viewing the streamers, the \emph{proportional indicator} strives directly to increase the viewing time of subscribers. Therefore, the latter indicator could lead to greater benefits for the platform in this new situation. In an average scenario where the platform aims to boost its success and profits (for instance, through advertising revenue) by increasing both the number of subscribers and the time they spend viewing the streamers, the platform should prioritize the \emph{subscriber-proportional indicator}.

\section{A case study: Twitch}

In the previous sections, we characterized and analyzed the \emph{uniform}, \emph{subscriber-uniform}, \emph{proportional}, and \emph{subscriber-proportional} indicators in the context of streaming platforms. In this section, we propose an application of these four indicators to measure the relevance of creators on the Twitch streaming platform Twitch.\footnote{https://www.twitch.tv/} 

We screened the 19 most-watched streamers worldwide during a three-week period from 23 December 2022 to 15 January 2023. In particular, we collected data on the viewers (identified through their nicknames) who were watching each streamer every streaming hour. In terms of the theoretical model introduced in Section \ref{model}, the services $N$ are the streamers, the subscribers $S$ are the viewers, and the consumption $C_{is}$ is the time each viewer spent watching the online content provided by each streamer. The subscription price is assumed to be fixed (and normalized to 1).\footnote{Information on the individual subscription fee that each viewer pays is not public. However, the platform has access to this information, which it can easily use to compute the distribution of the revenue that corresponds to each of the indicators.} 

The success of the Twitch platform is $\sigma=5,164,910$, which coincides with the number of viewers. Our database contains approximately five million unique users who watched one or more of the streamers. For the analyzed period, Table \ref{tabla1} shows the country, streaming language, number of viewers, overall share of viewers, exclusive viewers (those who exclusively consume the content of one streamer), and the percentage they represent out of the total amount of exclusive viewers. In terms of the number of users, AuronPlay, Ibai, and XQc are the three most relevant streamers, significantly outperforming the others. In terms of exclusive viewers, AuronPlay and XQc perform relatively well, but Ibai drops significantly.

\begin{center}
    [Insert Table \ref{tabla1} about here.]
\end{center}

\begin{table}
\begin{center}
\begin{tabular}{lllcccc}
\toprule
 &  &  & \multicolumn{2}{c}{Viewers} &  \multicolumn{2}{c}{Exclusive viewers} \\ \cmidrule(lr){4-5} \cmidrule(lr){6-7} 
Streamer & Country & Language & Number & \% & Number & \% \\ 
\midrule
Adin Ross & United States & English & 522425 & 6.97 & 279905 &7.49 \\
AuronPlay & Spain & Spanish & 758108 & 10.11 & 302403 & 8.09 \\
Casimito & Brazil & Portuguese & 224771 & 3.00 & 143199 & 3.83\\
ElSpreen & Argentina & Spanish  & 354882 & 4.73 & 160886 & 4.30 \\
elXokas & Spain & Spanish  & 237442 & 3.17 & 57847 & 1.55 \\
Fextralife & United States & English & 254907 & 3.40 & 228050 & 6.10 \\
Gaules  & Brazil & Portuguese & 239381 & 3.19 & 162742 & 4.35 \\
HasanAbi & United States & English & 372542 & 4.97 & 238815 & 6.39 \\
Ibai & Spain & Spanish & 738317 & 9.85 & 237515 & 6.35 \\
IlloJuan & Spain & Spanish & 476256 &  6.35 & 157036 & 4.20\\
Juansguarnizo & Mexico & Spanish & 400161 & 5.34 & 128466 & 3.44\\
Kai Cenat & United States & English & 501301 & 6.69 & 270001 & 7.22\\
Loltyler1 & United States & English & 299832 & 4.00 & 201687 & 5.40 \\
Loud$\_$coringa & Brazil & Portuguese & 279470 & 3.73 & 236475 & 6.33 \\
Roshtein & Malta & English & 4451 & 0.06 & 2782 & 0.07\\
Rubius & Spain & Spanish & 378760 &  5.05 & 115037 &3.08 \\
Tarik & United States & English & 410166 & 5.47 & 299966 & 8.02 \\
TheGrefg & Spain & Spanish & 318345 & 4.25 & 92534 & 2.48\\
XQc & Canada & English & 725958 & 9.68 & 422948 & 11.31\\ \hline
 &  &  & 7497475 & 100 & 3738294 & 100\\
\bottomrule
\end{tabular}
\caption{Viewers and exclusive viewers.\label{tabla1}}
\end{center}
\end{table}

Beyond the mere number of users or exclusive users, what is the specific impact of each streamer on the success of the Twitch? Table \ref{tabla2} illustrates the application of the \emph{uniform}, \emph{subscriber-uniform}, \emph{proportional}, and \emph{subscriber-proportional} indicators to our case study. For each metric, we indicate both the value and the percentage of success that is due to each streamer. Not surprisingly, for the \emph{uniform indicator} they are all equally relevant. Although this indicator may be seen as an uninformative measure, it serves as a useful benchmark for comparisons since, by definition, it coincides with the average share. The \emph{subscriber-uniform indicator} assigns the success of each individual viewer among the streamers whose content the subscriber viewed. Under this indicator, exclusive viewers have a significant impact on the distribution of success to each streamer. The subscription fee is fully assigned to that specific streamer. For the remaining viewers, success is distributed uniformly among all the streamers that they viewed. However, exclusive viewers do not completely determine the success rate. For example, Tarik is the third streamer with the most exclusive viewers, yet he is the sixth most successful streamer by this indicator.

The \emph{proportional indicator} (which implements the \emph{pro-rata} principle in \cite{Alaei22}) distributes the success proportionally to the overall consumption of each streamer. Notice that according to $R^P$, only a few streamers are above average (Adin Ross, AuronPlay, Ibai, IlloJuan, and XQc). Comparing the percentage of viewers with the \emph{proportional indicator}, we observe that the impact of Adin Ross is substantially higher than one might expect by focusing only on the users. His relevance increases in excess of 51\% to the $R^P$ indicator. IlloJuan, for example, is another streamer whose influence is undervalued. The other side of the coin is that the impact of TheGrefg is 35\% lower if we consider the \emph{proportional indicator}. 

The \emph{subscriber-proportional indicator} (which captures the idea behind the \emph{user-centric} principle in \cite{Alaei22}) assigns the success of each single viewer among the streamers whose content she has consumed. Therefore, by definition, the impact of an exclusive user is completely attributed to the unique streamer she watches. It is for this reason that the number of exclusive viewers has a higher impact in the \emph{subscriber-proportional indicator} than in the \emph{proportional indicator}. However, $R^{SP}$ and exclusive viewers are not completely correlated. For instance, Ibai is the seventh streamer in terms of the most exclusive viewers (6.35\%), but he rises to the third position for the \emph{subscriber-proportional indicator} (8.32\%), with an increase of a 31\%. This is because, in addition to his exclusive followers, his content is consumed by many other users.

We also observe several differences by comparing the \emph{proportional} and \emph{subscriber-proportional} indicators. The most significant change is the relevance of the streamer IlloJuan. He represents 10.15\% of the total success of the platform for the former, but only the 5.80\% for the latter. This is due to the fact that IlloJuan is one of the most watched streamers, relative to the total number of hours viewed, but the number of exclusive viewers (approximately $33\%$) is substantially lower than in other streamers with higher benefits. In contrast, Gaules is more relevant according to the \emph{subscriber-proportional indicator} compared to the \emph{proportional indicator}. The reason is that Gaules does not have many viewing hours compared to the total number of hours viewed by the users. However, this is a Portuguese streamer, while the others are Spanish or English, so he has a significant number of exclusive viewers, around $68\%$.

\begin{center}
    [Insert Table \ref{tabla2} about here.]
\end{center}

\begin{table}
\begin{center}
\begin{tabular}{lcccccccc}
\toprule
  & \multicolumn{2}{c}{$R^U$} &  \multicolumn{2}{c}{$R^{SU}$} & \multicolumn{2}{c}{$R^{P}$} & \multicolumn{2}{c}{$R^{SP}$}  \\ \cmidrule(lr){2-3} \cmidrule(lr){4-5} \cmidrule(lr){6-7}  \cmidrule(lr){8-9}
 Streamer & Value & \% &  Value & \% &  Value & \% &  Value & \%\\
\midrule
Adin Ross       & 271837.4 & 5.26 & 384314.07 & 7.44 & 545333.40 & 10.56 & 399648.12 & 7.74  \\
AuronPlay       & 271837.4 & 5.26 & 467563.86 & 9.05 & 448855.80 & 8.69  & 462333.00 & 8.95  \\
Casimito        & 271837.4 & 5.26 & 181757.52 & 3.52 & 219242.70 & 4.24  & 193421.99 & 3.74  \\
ElSpreen        & 271837.4 & 5.26 & 228803.63 & 4.43 & 189751.10 & 3.67  & 228175.61 & 4.42  \\
elXokas         & 271837.4 & 5.26 & 117556.95 & 2.28 & 106974.10 & 2.07  & 100438.95 & 1.94  \\
Fextralife      & 271837.4 & 5.26 & 239775.26 & 4.64 & 172821.40 & 3.35  & 238549.53 & 4.62  \\
Gaules          & 271837.4 & 5.26 & 198698.27 & 3.85 & 105343.10 & 2.04  & 185148.12 & 3.58  \\
HasanAbi        & 271837.4 & 5.26 & 294519.76 & 5.70 & 191565.80 & 3.71  & 277481.56 & 5.37 \\
Ibai            & 271837.4 & 5.26 & 417746.54 & 8.09 & 466978.00 & 9.04  & 429475.25 & 8.32 \\
IlloJuan        & 271837.4 & 5.26 & 269565.17 & 5.22 & 524492.80 & 10.15 & 299687.49 & 5.80  \\
Juansguarnizo   & 271837.4 & 5.26 & 223595.42 & 4.33 & 255009.10 & 4.94  & 222511.11 & 4.31 \\
Kai Cenat       & 271837.4 & 5.26 & 369516.65 & 7.15 & 266601.80 & 5.16  & 353599.40 & 6.85  \\
Loltyler1       & 271837.4 & 5.26 & 243739.64 & 4.72 & 188135.50 & 3.64  & 244417.87 & 4.73  \\
Loud$\_$coringa & 271837.4 & 5.26 & 256230.97 & 4.96 & 252097.30 & 4.88  & 258610.57 & 5.00  \\
Roshtein        & 271837.4 & 5.26 & 3409.45 & 0.07 & 5992.60   & 0.11  & 3537.18   & 0.06  \\
Rubius          & 271837.4 & 5.26 & 204089.17 & 3.95 & 253755.40 & 4.91  & 198031.44 & 3.83  \\
Tarik           & 271837.4 & 5.26 & 346841.69 & 6.72 & 219131.10 & 4.24  & 351297.07 & 6.80  \\
TheGrefg        & 271837.4 & 5.26 & 166702.99 & 3.23 & 141868.60 & 2.75  & 160103.74 & 3.10  \\
XQc             & 271837.4 & 5.26 & 550482.98 & 10.66 & 610960.50 & 11.83 & 558442.00 & 10.81 \\
\bottomrule
\end{tabular}
\caption{Relevance of streamers according to the proposed indicators.\label{tabla2}}
\end{center}
\end{table}

At this point, a question that arises is how the relevance of each streamer can be translated into revenue allocation. One way to do this is to distribute the part of the total revenue from subscription fees that the platform dedicates to rewarding streamers in proportion to the relevance of the platform's success that the indicators attribute to them. Note that this form of distribution will not modify in any way the structure of the relevance indicators presented, since the percentages obtained in Table \ref{tabla2} have been obtained using these indicators. In this sense, the comparative analysis carried out in the previous paragraphs on the indicators would also be valid for the allocation of revenue among streamers in proportion to these indicators.

\section{Concluding remarks}

In recent years, streaming services have risen significantly in popularity. In return for a subscription, consumers have unlimited access to the catalog of services that platforms offers, such as movies, TV shows, music, and books. Evidently, not all services are equally relevant for attracting users' interest to pay for a subscription. In this paper, we have presented a model to measure the relevance of each service with respect to the global success of the platform. We have addressed this problem using an axiomatic approach, and we have provided normative foundations for four basic indicators: the \emph{uniform}, \emph{subscriber-uniform}, \emph{proportional}, and \emph{subscriber-proportional} indicators. These four metrics are characterized by means of several axioms, which represent different principles of fairness and stability. Table \ref{tabla2} summarizes the properties fulfilled by each indicator and the ones that are necessary for the characterization. Moreover, different issues related to information, structure of the indicators, and axioms have been analyzed to establish their importance in the context of streaming platforms. This analysis also includes an understanding of which relevance indicators may be preferred by streamers and the marginal impact on a streamer's relevance when viewing time increases (or decreases). We have also discussed the question of which indicators are the most interesting for the platform.

In the particular context of music streaming platforms and the distribution of revenues among artists, two mechanisms have emerged as focal allocation rules: the \emph{pro-rata} and the \emph{user-centric} rules (see \cite{Meyn23}). In our model, the \emph{pro-rata} and \emph{user-centric} schemes coincide with the \emph{proportional} and \emph{subscriber-proportional} indicators, respectively. Therefore, Theorems \ref{thm1} and \ref{thm2} also reveal characterizations of these two allocation rules. In other words, based on principles of fairness, stability, and non-manipulability, we provide normative foundations for the two most prominent revenue allocation mechanisms in the music streaming market.  

To illustrate the performance of the four indicators, we have analyzed the streaming platform Twitch. Over a three-week period, we recorded the number of viewers and viewing times of the 19 most-followed Twitch streamers worldwide. Based on the normative approach presented in Section 4, we have concluded that the mere number of viewers is not a suitable metric to capture the impact of each streamer. We have also found that exclusive viewers (i.e., users who consume the content of a unique streamer) have a higher influence in the \emph{subscriber-proportional indicator} than in the \emph{proportional indicator}.

We acknowledge that there are still some issues that deserve a deeper analysis. This paper focuses on measuring the influence of each service from the point of view of consumption. However, there are cases where the perspective of the production should also be considered. For instance, considering TV series, the viewers' reaction is important, but the production cost cannot be ignored. In this respect, a natural extension of the proposed model is to define indicators that consider both parts of the market. The literature on resource allocation and attribution problems is quite extensive, and other metrics and axioms can also be explored.

\newpage

\begin{thebibliography}{21}
    \providecommand{\natexlab}[1]{#1}
    \providecommand{\url}[1]{\texttt{#1}}
    \providecommand{\urlprefix}{URL }
    
    \bibitem[{Adams and Yellen(1976)}]{Adams76}
    Adams, W.~J. and Yellen, J.~L. (1976).
    \newblock Commodity bundling and the burden of monopoly.
    \newblock \emph{The Quarterly Journal of Economics}, 90:475--497.
    
    \bibitem[{Alaei et~al.(2022)Alaei, Makhdoumi, Malekian, and
      Peke\u{c}}]{Alaei22}
    Alaei, S., Makhdoumi, A., Malekian, A., and Peke\u{c}, S. (2022).
    \newblock Revenue-{Sharing} {Allocation} {Strategies} for {Two}-{Sided} {Media}
      {Platforms}: {Pro}-{Rata} vs. {User}-{Centric}.
    \newblock \emph{Management Science}, 68:8699--8721.
    
    \bibitem[{Albino et~al.(2008)Albino, Scaruffi, Moretti, Coco, Truini,
      Cristofano, Cavazzana, Stigliani, Bonassi, and Tonini}]{Albino08}
    Albino, D., Scaruffi, P., Moretti, S., Coco, S., Truini, M., Cristofano, C.~D.,
      Cavazzana, A., Stigliani, S., Bonassi, S., and Tonini, G.~P. (2008).
    \newblock Identification of low intratumoral gene expression heterogeneity in
      neuroblastic tumors by genome-wide expression analysis and game theory.
    \newblock \emph{Cancer}, 113:1412--1422.

    \bibitem[{Algaba et~al.(2019b)Fragnelli Llorca, and S\'anchez-Soriano}]{Algaba19b}
    Algaba, E., Fragnelli, V., Llorca, N., and S\'anchez-Soriano, J. (2019).
    \newblock Horizontal cooperation in a multimodal public transport system: The
profit allocation problem.
    \newblock \emph{European Journal of Operational Research}, 275:659--665..


  \bibitem[{Algaba et~al.(2019)Fragnelli and S\'anchez-Soriano}]{Algaba19}
    Algaba, E., Fragnelli, V. and S\'anchez-Soriano, J. (2022).
    \newblock Handbook of the Shapley value.
    \newblock \emph{CRC Press}.
    
    \bibitem[{Berganti{\~{n}}os and Moreno-Ternero(2015)}]{Bergantinos15}
    Berganti{\~{n}}os, G. and Moreno-Ternero, J.~D. (2015).
    \newblock The axiomatic approach to the problem of sharing the revenue from
      museum passes.
    \newblock \emph{Games and Economic Behavior}, 89:78--92.
    
    \bibitem[{Berganti{\~n}os and Moreno-Ternero(2023)}]{Bergantinos23}
    ---{}---{}--- (2023).
    \newblock An axiomatic approach to the revenue sharing in the streaming
      industry.
    \newblock \emph{22nd annual SAET Conference Paris, France}.
    
    \bibitem[{Brander et~al.(2011)Brander, Bruno, Hobday, and Schoeman}]{Brander11}
    Brander, K., Bruno, J., Hobday, A., and Schoeman, D. (2011).
    \newblock The value of attribution.
    \newblock \emph{Nature Climate Change}, 1:70--71.
    
    \bibitem[{Burger et~al.(2020)Burger, Wentz, and Horton}]{Burger20}
    Burger, M., Wentz, J., and Horton, R. (2020).
    \newblock The law and science of climate change attribution.
    \newblock \emph{Columbia Journal of Environmental Law}, 5:57--240.
    
    \bibitem[{Dimont(2017)}]{Dimont17}
    Dimont, J. (2017).
    \newblock Royalty inequity: {W}hy music streaming services should switch to a
      per-subscriber model.
    \newblock \emph{CU Hastings Law Journal}, 69:675--700.
    
    \bibitem[{Ginsburgh and Zang(2003)}]{Ginsburgh03}
    Ginsburgh, V. and Zang, I. (2003).
    \newblock The museum pass game and its value.
    \newblock \emph{Games and Economic Behavior}, 43:322--325.
    
    \bibitem[{Jari(2018)}]{Prorata18}
    Jari, M. (2018).
    \newblock Pro-rata and user-centric distribution models: {A} comparative study.
    \newblock Technical report.
    
    \bibitem[{Ju et~al.(2007)Ju, Miyagawa, and Sakai}]{Ju07}
    Ju, B.~G., Miyagawa, E., and Sakai, T. (2007).
    \newblock {N}on-manipulable division rules in claim problems and
      generalizations.
    \newblock \emph{Journal of Economic Theory}, 132:1--16.

    \bibitem[{L\'opez-Navarrete et~al.(2019)L\'opez-Navarrete, S\'anchez-Soriano, and Bonastre}]{Lopez19}
    L\'opez-Navarrete, F., S\'anchez-Soriano, J., and Bonastre, O.~M. (2019).
    \newblock Allocating revenues in a Smart TV ecosystem.
    \newblock \emph{International Transactions in Operational Research}, 26:1611--1632.

    \bibitem[{L\'opez-Navarrete et~al.(2023)L\'opez-Navarrete, S\'anchez-Soriano, and Bonastre}]{Lopez23}
    ---{}---{}--- (2023).
    \newblock Dynamic generation and attribution of revenues in a video platform.
    \newblock ArXiv, 4859841 [math.OC].
    
    \bibitem[{Lucchetti et~al.(2010)Lucchetti, Moretti, Patrone, and
      Radrizzani}]{Lucchetti10}
    Lucchetti, R., Moretti, S., Patrone, F., and Radrizzani, P. (2010).
    \newblock The {S}hapley and {B}anzhaf values in microarray games.
    \newblock \emph{Computers {\&} Operations Research}, 37:1406--1412.
    
    \bibitem[{Manshadi et~al.(2023)Manshadi, Niazadeh, and Rodilitz}]{Manshadi23}
    Manshadi, V., Niazadeh, R., and Rodilitz, S. (2023).
    \newblock Fair dynamic rationing.
    \newblock \emph{Management Science}, forthcoming.
    
    \bibitem[{Mart\'inez and Moreno-Ternero(2022)}]{Martinez22d}
    Mart\'inez, R. and Moreno-Ternero, J.~D. (2022).
    \newblock An axiomatic approach towards pandemic performance indicators.
    \newblock \emph{Economic Modelling}, 116:105983.
    
    \bibitem[{Mart{\'{\i}}nez and S{\'{a}}nchez-Soriano(2021)}]{Martinez21}
    Mart{\'{\i}}nez, R. and S{\'{a}}nchez-Soriano, J. (2021).
    \newblock Mathematical indices for the influence of risk factors on the
      lethality of a disease.
    \newblock \emph{Journal of Mathematical Biology}, 83:74.

    \bibitem[{Mart{\'{\i}}nez and S{\'{a}}nchez-Soriano(2023)}]{Martinez23}
    Mart{\'{\i}}nez, R. and S{\'{a}}nchez-Soriano, J. (2023).
    \newblock Order preservation with dummies in the museum pass problem.
    \newblock \emph{Annals of Operations Research}, forthcoming.
    
    \bibitem[{Mart{\'{\i}}nez et~al.(2022)Mart{\'{\i}}nez, S{\'{a}}nchez-Soriano,
      and Llorca}]{Martinez22b}
    Mart{\'{\i}}nez, R., S{\'{a}}nchez-Soriano, J., and Llorca, N. (2022).
    \newblock Assessments in public procurement procedures.
    \newblock \emph{Omega}, 111:102660.
    
    \bibitem[{Meyn et~al.(2023)Meyn, Kandziora, Albers, and Clement}]{Meyn23}
    Meyn, J., Kandziora, M., Albers, S., and Clement, M. (2023).
    \newblock Consequences of platforms' remuneration models for digital content:
      initial evidence and a research agenda for streaming services.
    \newblock \emph{Journal of the Academy of Marketing Science}, 51:114--131.
    
    \bibitem[{Moretti et~al.(2007)Moretti, Patrone, and Bonassi}]{Moretti07}
    Moretti, S., Patrone, F., and Bonassi, S. (2007).
    \newblock The class of microarray games and the relevance index for genes.
    \newblock \emph{{TOP}}, 15:256--280.

     \bibitem[{O'Neill(1982)}]{Oneill82}
    O'Neill, B. (1982).
    \newblock A problem of rights arbitration from the Talmud.
    \newblock \emph{{Mathematical social sciences}}, 2:345--371.
    
    \bibitem[{Page and Safir(2018{\natexlab{a}})}]{Page18b}
    Page, W. and Safir, D. (2018{\natexlab{a}}).
    \newblock Money in, money out: lessons from cmos in allocating and distributing
      licensing revenue.
    \newblock Technical report.
    
    \bibitem[{Page and Safir(2018{\natexlab{b}})}]{Page18}
    ---{}---{}--- (2018{\natexlab{b}}).
    \newblock User-centric revisited: The unintended consequences ofroyalty
      distribution.
    \newblock Technical report.

    \bibitem[{Roth(1988)}]{Roth88}
    Roth, A. E. (1988).
    \newblock The Shapley value: essays in honor of Lloyd S. Shapley.
    \newblock \emph{Cambridge University Press}.

    \bibitem[{Shapley(1953))}]{Shapley53}
    Shapley, L.S. (1953).
    \newblock A value for n-person games.
    \newblock \emph{Contributions to the Theory of Games}, 2:307--317.
       
    
    \bibitem[{Singal et~al.(2022)Singal, Besbes, Desir, Goyal, and
      Iyengar}]{Singal22}
    Singal, R., Besbes, O., Desir, A., Goyal, V., and Iyengar, G. (2022).
    \newblock Shapley meets uniform: {A}n axiomatic framework for attribution in
      online advertising.
    \newblock \emph{Management Science}, 68:7457--7479.
    
    \end{thebibliography}

\newpage
\section*{Appendix A}
Proof of properties in Table \ref{Table1}.

The uniform indicator satisfies \emph{strong symmetry}. Let $(N,S,p,C) \in \mathcal{D}$. Let $i,j \in N$ such that $C_{is}\cdot C_{js}\neq 0$  or $C_{is}=C_{js}=0$ for all $s\in S$. Then we have $R^{U}_i(N,S,p,C)=\frac{\sigma}{n}=R^{U}_j(N,S,p,C)$.

The uniform indicator satisfies \emph{composition}. Let $(N,S,p,C), (N,S',p',C') \in \mathcal{D}$ such that $S \cap S' = \emptyset$. Let $i \in N$. Then we have that
   \begin{align*}
			R_i^{U}(N,S \cup S',p \oplus p' ,C \oplus C') &= \sum_{s \in S \cup S'} \dfrac{(p \oplus p')_s}{|N|}  \\
			&= \sum_{s \in S} \dfrac{p_s}{|N|}  + \sum_{s \in S'} \dfrac{p'_s}{|N|}   \\
			&= R_i^{U}(N,S,p,C) + R_i^{U}(N,S',p',C').
	\end{align*}

The uniform indicator satisfies \emph{sharing proofness}. Let $(N,S,p,C) \in \mathcal{D}$. Let $S' \subseteq S$ and $s \in S'$ such that $p'_s=\sum_{t\in S'}p_t$ and $C'_{is}=\sum_{t \in S'} C_{it}$ for all $i\in N$ and $C'_{is}=C_{is}$ for all $i\in N$ and $s\in S\backslash S'$. Then
\begin{align*}
    R^U_i \left( N ,\{s\}\cup S\backslash S',(p'_s,p_{S\backslash S'}),C' \right) &= \dfrac{p'_s+\sum_{t\in S\backslash S'}p_t}{ |N|}  \\
    &=\dfrac{\sum_{t\in S'}p_t+\sum_{t\in S\backslash S'}p_t}{ |N|}\\
    &=\dfrac{\sum_{t\in S}p_t}{ |N|}\\
    &= R^{U}_i(N,S,p,C).
    \end{align*}

The uniform indicator does not satisfy \emph{non-manipulabity}. Consider a platform where $N=\{1,2,3\}$, $S=\{1\}$, $p_1=1$, and $C$ is given by
$$
\left(
\begin{array}{cccccc}
	0  \\
	1  \\
	2.5  \\
\end{array}
\right).
$$

The uniform indicator is equal to $R^{U}_i(N,S,p,C)=\left(\frac{1}{3},\frac{1}{3},\frac{1}{3}\right)$. Now, consider $N'=\{1,2\}$ and $C'_{11}=C_{11}+C_{12}=1$. Then
$$
R_1 \left(\{1\} \cup N \backslash N',S,p,(C'_{1 \cdot},C_{N \backslash N'}) \right)=\frac{1}{2} \neq \sum_{j \in N'} R_j(N,S,p,C)=\frac{2}{3}.
$$

The uniform indicator does not satisfy \emph{nullity}. Let $(N,S,p,C) \in \mathcal{D}$. Let $i \in N$ such that $C_{is}=0$ for all $s\in S$. Then
$$R^{U}_i(N,S,p,C)=\frac{\sigma}{|N|}\neq 0.$$

The subscriber-uniform indicator satisfies \emph{symmetry}. Let $(N,S,p,C) \in \mathcal{D}$. Let $i,j \in N$ such that $C_{is}=C_{js}$ for all $s\in S$. Then $i\in N_s$ if and only if $j\in N_s$. Therefore, we have $R^{SU}_i(N,S,p,C)=R^{SU}_j(N,S,p,C)$.

The subscriber-uniform satisfies \emph{homogeneity}.
Let $(N,S,p,C) \in \mathcal{D}$ and $\lambda \in  \mathbb{R}_{++}$. Let $i\in N$. We have that
 \begin{align*}
			R_i^{SU}(N,S,p,\lambda C) &=\sum_{s \in S, i\in N_s} \dfrac{1}{\left| N_s\right|}p_s=R_i^{P}(N,S,p, C),
	\end{align*}
where $ N_s=\left\{ j \in N: C_{js} \neq 0\right\}$.

The subscriber-uniform indicator does not satisfy \emph{consumption sensitivity}. Let us consider the following two successions of consumption matrices:
    $$
    C^{1n}=\left(
    \begin{array}{c}
     \frac{1}{n}\\
     \frac{1}{n}\\
     \frac{1}{n}\\
    \end{array}
    \right)
    \text{ and }
    C^{2n}=\left(
    \begin{array}{c}
     0\\
     \frac{1}{n}\\
     \frac{2}{n}\\
    \end{array}
    \right).
    $$
    It is obvious that $\|C^{1n} - C^{2n}\|$ goes to $0$ when $n$ goes to infinity, but $\|R^{SU}(N,S,p,C^{1n})- R^{SU}(N,S,p,C^{2n})\|=\|(\frac{p}{3},\frac{p}{3},\frac{p}{3}) - (0,\frac{p}{2},\frac{p}{2})\| = \frac{2p}{3} >> 0$ for all $n$.

The subscriber-uniform indicator satisfies \emph{composition}. Let $(N,S,p,C), (N,S',p',C') \in \mathcal{D}$ such that $S \cap S' = \emptyset$. Let $i \in N$. We have that
   \begin{align*}
			R_i^{SU}(N,S \cup S',p \oplus p' ,C \oplus C') &= \sum_{s \in S \cup S'} \dfrac{1}{\left| N_s\right|} (p \oplus p')_s  \\
			&= \sum_{s \in S}\dfrac{1}{\left| N_s\right|} p_s  + \sum_{s \in S'} \dfrac{1}{\left| N_s\right|}p'_s  \\
			&= R_i^{SU}(N,S,p,C) + R_i^{SU}(N,S',p',C').
	\end{align*}

The subscriber-uniform indicator does not satisfy \emph{sharing proofness}. Consider a platform where $N=\{1,2,3\}$, $S=\{1,2\}$, $p=(1,1)$, and $C$ is given by
 $$
\left(
\begin{array}{cccccc}
	0 & 2\\
	1 & 0\\
	1 & 0\\
\end{array}
\right).
$$

Then $R^{SU}(N,S,p, C)=\left(1,\frac{1}{2},\frac{1}{2}\right)$. Let $S'=\{1,2\} \subseteq S$ and $s=1 \in S'$ such that $p'_s=\sum_{t\in S'}p_t$ and $C'_{is}=\sum_{t \in S'} C_{it}$ for all $i\in N$. Then $R^{SP}(N,S',p', C')=\left(\frac{2}{3},\frac{2}{3},\frac{2}{3}\right)\neq R^{SP}(N,S,p, C)$.

The proportional indicator satisfies \emph{symmetry}. Let $(N,S,p,C) \in \mathcal{D}$. Let $i,j \in N$ such that $C_{is}=C_{js}$ for all $s\in S$. We have that
   \begin{align*}
			R_i^{P}(N,S,p,C) &=\frac{\|C_{i \cdot}\|}{\displaystyle\sum_{t \in N} \|C_{t \cdot}\|} \sigma=\frac{\sum_{s\in S}C_{is}}{\displaystyle\sum_{t \in N} \|C_{t \cdot}\|} \sigma=\frac{\sum_{s\in S}C_{js}}{\displaystyle\sum_{t \in N} \|C_{t \cdot}\|} \sigma=\frac{\|C_{j \cdot}\|}{\displaystyle\sum_{t \in N} \|C_{t \cdot}\|} \sigma=R_j^{P}(N,S,p,C).
	\end{align*}

The proportional indicator does not satisfy \emph{strong symmetry}. Consider a platform where $N=\{1,2,3\}$, $S=\{1\}$, $p_1$, where $C$ is given by
 $$
\left(
\begin{array}{cccccc}
	0  \\
	1  \\
	2  \\
\end{array}
\right).
$$
Then $C_{21}\cdot C_{32}\neq 0$ and $R_2^P(N,S,p,C)=\frac{p_1}{3}\neq \frac{2p_1}{3}=R_3^P(N,S,p,C)$.

The proportional indicator satisfies \emph{homogeneity}.
Let $(N,S,p,C) \in \mathcal{D}$ and $\lambda \in  \mathbb{R}_{++}$.  Let $i\in N$. We have that
 \begin{align*}
			R_i^{P}(N,S,p,\lambda C) &=\frac{\|\lambda C_{i \cdot}\|}{\displaystyle\sum_{j \in N} \|\lambda C_{j \cdot}\|} \sigma=\frac{\lambda\| C_{i \cdot}\|}{\displaystyle\sum_{j \in N} \lambda\| C_{j \cdot}\|} \sigma=R_i^{P}(N,S,p, C) .
	\end{align*}

The proportional indicator does not satisfy \emph{composition}. Consider a platform where $N=\{1,2,3\}$, $S=\{1\}$, $S'=\{2\}$, $p_1=1$, where $C^1$ and $C^2$ are given by
 $$
\left(
\begin{array}{cccccc}
	0  \\
	1  \\
	1  \\
\end{array}
\right),
$$
$$
\left(
\begin{array}{cccccc}
	2  \\
	1  \\
	1  \\
\end{array}
\right).
$$

Then $R^{P}(N,\{1\},p_1, C^1)=\left(0,\frac{1}{2},\frac{1}{2}\right)$ and $R^{P}(N,\{2\},p_2, C^2)=\left(\frac{1}{2},\frac{1}{4},\frac{1}{4}\right)$. By another way, $R^{P}(N,\{1\} \cup \{2\},p_1 \oplus p_2 ,C^1 \oplus C^2)=\left(\frac{2}{3},\frac{2}{3},\frac{2}{3}\right)\neq \left(\frac{1}{2},\frac{3}{4},\frac{3}{4}\right)=R^{P}(N,\{1\},p_1, C^1)+R^{P}(N,\{2\},p_2, C^2)$.

The proportional indicator satisfies \emph{nullity}. Let $(N,S,p,C) \in \mathcal{D}$. Let $i \in N$ such that $C_{is}=0$ for all $s\in S$. Then
$$R^{P}_i(N,S,p,C)=\frac{\| C_{i \cdot}\|}{\displaystyle\sum_{j \in N} \| C_{j \cdot}\|} \sigma= \frac{\sum_{s\in S}C_{is}}{\displaystyle\sum_{j \in N} \| C_{j \cdot}\|} \sigma=0.$$

The subscriber-proportional indicator satisfies \emph{symmetry}. Let $(N,S,p,C) \in \mathcal{D}$. Let $i,j \in N$ such that $C_{is}=C_{js}$ for all $s\in S$. We have that
   \begin{align*}
			R_i^{SP}(N,S,p,C) &= \sum_{s \in S} \dfrac{C_{is}}{\|C_{\cdot s}\|} p_s= \sum_{s \in S} \dfrac{C_{js}}{\|C_{\cdot s}\|} p_s=R_j^{SP}(N,S,p,C).
	\end{align*}

The subscriber-proportional indicator does not satisfy \emph{strong symmetry}. Consider a platform where $N=\{1,2,3\}$, $S=\{1\}$, $p_1$, where $C$ is given by
 $$
\left(
\begin{array}{cccccc}
	0  \\
	1  \\
	2  \\
\end{array}
\right).
$$
Then $C_{21}\cdot C_{32}\neq 0$ and $R_2^{SP}(N,S,p,C)=\frac{p_1}{3}\neq \frac{2p_1}{3}=R_3^{SP}(N,S,p,C)$.

The subscriber-proportional indicator satisfies \emph{homogeneity}. Let $(N,S,p,C) \in \mathcal{D}$ and $\lambda \in  \mathbb{R}_{++}$. Let $i\in N$. We have that
 \begin{align*}
			R_i^{SP}(N,S,p,\lambda C) &=\sum_{s \in S} \dfrac{\lambda C_{is}}{\|\lambda C_{\cdot s}\|} p_s=\sum_{s \in S} \dfrac{\lambda C_{is}}{\lambda \| C_{\cdot s}\|} p_s=R_i^{SP}(N,S,p, C) .
	\end{align*}

The subscriber-proportional indicator does not satisfy \emph{consumption sensitivity}. Let us consider the following two successions of consumption matrices:
    $$
    C^{1n}=\left(
    \begin{array}{c}
     \frac{1}{n}\\
     \frac{1}{n}\\
     \frac{1}{n}\\
    \end{array}
    \right)
    \text{ and }
    C^{2n}=\left(
    \begin{array}{c}
     0\\
     \frac{1}{n}\\
     \frac{2}{n}\\
    \end{array}
    \right).
    $$
    It is obvious that $\|C^{1n} - C^{2n}\|$ goes to $0$ when $n$ goes to infinity, but $\|R^{SP}(N,S,p,C^{1n})- R^{SP}(N,S,p,C^{2n})\|=\|(\frac{p}{3},\frac{p}{3},\frac{p}{3}) - (0,\frac{p}{3},\frac{2p}{3})\| = \frac{2p}{3} >> 0$ for all $n$.

The subscriber-proportional indicator does not satisfy \emph{sharing proofness}. Consider a platform where $N=\{1,2,3\}$, $S=\{1,2\}$, $p=(1,1)$, and $C$ is given by
 $$
\left(
\begin{array}{cccccc}
	0 & 2\\
	1 & 1\\
	1 & 1\\
\end{array}
\right)
$$

Then, $R^{SP}(N,S,p, C)=\left(\frac{1}{2},\frac{3}{4},\frac{3}{4}\right)$. Let $S'=\{1,2\} \subseteq S$ and $s=1 \in S'$ such that $p'_s=\sum_{t\in S'}p_t$ and $C'_{is}=\sum_{t \in S'} C_{it}$ for all $i\in N$, then $R^{SP}(N,S',p', C')=\left(\frac{2}{3},\frac{2}{3},\frac{2}{3}\right)\neq R^{SP}(N,S,p, C)$.

The subscriber-proportional indicator satisfies \emph{nullity}.  Let $(N,S,p,C) \in \mathcal{D}$. Let $i \in N$ such that $C_{is}=0$ for all $s\in S$, then
$$R^{SP}_i(N,S,p,C)=\sum_{s \in S} \dfrac{ C_{is}}{\| C_{\cdot s}\|} p_s=0.$$

\section{Appendix B}

Proof of Lemma \ref{le.0.}.

\begin{proof}
    Let $(N,S,p,C) \in \mathcal{D}$, and suppose we have a streamer $i\in N$ such that $$\displaystyle\sum_{s\in S, i\in N_s}\frac{\left(|N|-|N_s|\right)}{|N_s|}p_s\geq \sum_{s\in S, i\notin N_s}p_s.$$
    Then we have that
       \begin{align*}
			 R_i^{U}(N,S,p  ,C)&= \sum_{s \in S}  \frac{1}{|N|}p_s=\sum_{s \in S, i\in N_s}  \frac{1}{|N|}p_s+\sum_{s \in S, i\notin N_s}  \frac{1}{|N|}p_s\\
    &\leq \sum_{s \in S, i\in N_s}  \frac{1}{|N|}p_s+\sum_{s\in S, i\in N_s}\frac{\left(|N|-|N_s|\right)}{|N||N_s|}p_s\\
    &=\sum_{s \in S, i\in N_s}\frac{\left(|N_s|+|N|-|N_s|\right)}{|N||N_s|}p_s\\
    &=\sum_{s \in S, i\in N_s}\frac{1}{|N_s|}p_s=R_i^{SU}(N,S,p  ,C).\\
	\end{align*}
\end{proof}

Proof of Lemma \ref{le.1.}.

\begin{proof}
    Let $(N,S,p,C) \in \mathcal{D}$, and suppose we have a streamer $i\in N$ such that $C_{is}>0$ for all $s\in S$. Then we have that
       \begin{align*}
			 R_i^{SU}(N,S,p  ,C)&= \sum_{s \in S, i\in N_s}\frac{1}{|N_s|}p_s=\sum_{s \in S}\frac{1}{|N_s|}p_s\geq \sum_{s \in S}  \frac{1}{|N|}p_s=R_i^{SU}(N,S,p  ,C).\\
	\end{align*}
\end{proof}

Proof of Lemma \ref{le.2.}.

\begin{proof}
    Let $(N,S,p,C) \in \mathcal{D}$, and suppose we have a streamer $i\in N$  such that $\|C_{i\cdot}\|\geq \frac{\sum_{j\in N\backslash\{i\}}\|C_{j\cdot}\|}{|N|-1}$. Then we have that
        \begin{align*}
    R^P_i(N,S,p,C) &= \frac{\|C_{i \cdot}\|}{\displaystyle\sum_{j \in N} \|C_{j \cdot}\|} \sigma\geq \frac{\frac{\displaystyle\sum_{j\in N\backslash\{i\}}\|C_{j\cdot}\|}{|N|-1}}{\displaystyle\sum_{j \in N} \|C_{j \cdot}\|}\sigma=\frac{\displaystyle\sum_{j\in N\backslash\{i\}}\|C_{j\cdot}\|}{\left(|N|-1\right)\displaystyle\sum_{j \in N} \|C_{j \cdot}\|}\sigma\\
    &=\frac{\displaystyle\sum_{j\in N\backslash\{i\}}\|C_{j\cdot}\|}{\left(|N|-1\right)\displaystyle\left(\sum_{j \in N\backslash\{i\}} \|C_{j \cdot}\|+\|C_{i \cdot}\|\right)}\sigma\\
    &\geq \frac{\displaystyle\sum_{j\in N\backslash\{i\}}\|C_{j\cdot}\|}{\left(|N|-1\right)\displaystyle\left(\sum_{j \in N\backslash\{i\}} \|C_{j \cdot}\|+\frac{\sum_{j\in N\backslash\{i\}}\|C_{j\cdot}\|}{|N|-1}\right)}\sigma\\
    &= \frac{\displaystyle\sum_{j\in N\backslash\{i\}}\|C_{j\cdot}\|}{\left(|N|-1\right)\displaystyle\left(\frac{|N|}{|N|-1}\displaystyle\sum_{j \in N\backslash\{i\}} \|C_{j \cdot}\|\right)}\sigma\\
    &=\frac{\displaystyle\sum_{j\in N\backslash\{i\}}\|C_{j\cdot}\|}{\displaystyle|N|\sum_{j \in N\backslash\{i\}} \|C_{j \cdot}\|}\sigma=\frac{\sigma}{|N|}= R_i^{U}(N,S,p  ,C).
	\end{align*}
\end{proof}

Proof of Lemma \ref{le.3.}.

\begin{proof}
    Let $(N,S,p,C) \in \mathcal{D}$, and suppose we have a streamer $i\in N$  such that $C_{is}\geq \frac{\displaystyle\sum_{j\in N\backslash\{i\}}C_{js}}{|N|-1}$ for all $s\in S$. Then we have that
        \begin{align*}
        R^{SP}_i(N,S,p,C) &= \sum_{s \in S} \dfrac{C_{is}}{\|C_{\cdot s}\|} p_s= \sum_{s \in S} \dfrac{C_{is}}{\displaystyle\sum_{j\in N\backslash\{i\}}C_{js} +C_{is}} p_s\\
        &\geq \sum_{s \in S} \dfrac{\frac{\displaystyle\sum_{j\in N\backslash\{i\}}C_{js}}{|N|-1}}{\displaystyle\sum_{j\in N\backslash\{i\}}C_{js} +\frac{\displaystyle\sum_{j\in N\backslash\{i\}}C_{js}}{|N|-1}} p_s\\
        &= \sum_{s \in S} \dfrac{\frac{\displaystyle\sum_{j\in N\backslash\{i\}}C_{js}}{|N|-1}}{\frac{|N|}{|N|-1}\displaystyle\sum_{j\in N\backslash\{i\}}C_{js}} p_s= \sum_{s \in S}  \frac{1}{|N|}p_s=R_i^{SU}(N,S,p  ,C).\\
	\end{align*}
\end{proof}

Proof of Lemma \ref{le.4.}.

\begin{proof}
    Let $(N,S,p,C) \in \mathcal{D}$, and suppose we have a streamer $i\in N$  such that $C_{is}\geq \frac{\sum_{j\in N_s\backslash\{i\}}C_{js}}{|N_s|-1}$ for all $s\in S$ with $i\in N_s$. Then we have that
        \begin{align*}
        R^{SP}_i(N,S,p,C) &= \sum_{s \in S} \dfrac{C_{is}}{\|C_{\cdot s}\|} p_s=\sum_{s \in S, i\in N_s} \dfrac{C_{is}}{\|C_{\cdot s}\|} p_s =\sum_{s \in S, i\in N_s} \dfrac{C_{is}}{\displaystyle\sum_{j\in N\backslash\{i\}}C_{js} +C_{is}} p_s\\
        &\geq \sum_{s \in S, i\in N_s} \dfrac{\frac{\displaystyle\sum_{j\in N_s\backslash\{i\}}C_{js}}{|N_s|-1}}{\displaystyle\sum_{j\in N\backslash\{i\}}C_{js} +\frac{\displaystyle\sum_{j\in N_s\backslash\{i\}}C_{js}}{|N_s|-1}} p_s\\
        &= \sum_{s \in S, i\in N_s} \dfrac{\frac{\displaystyle\sum_{j\in N_s\backslash\{i\}}C_{js}}{|N_s|-1}}{\frac{|N_s|}{|N_s|-1}\displaystyle\sum_{j\in N_s\backslash\{i\}}C_{js}} p_s=\sum_{s \in S, i\in N_s}\frac{1}{|N_s|}p_s=R_i^{SU}(N,S,p  ,C).\\
	\end{align*}
\end{proof}

Proof of Lemma \ref{le.6.}.

\begin{proof}
   Let $(N,S,p,C), (N',S,p,C') \in \mathcal{D}$, let $i\in N$ where $N'=N\backslash\{i\}\cup\{i'_1,\hdots,i'_m\}$ with $m\geq 2$ and $C_{is}=\sum_{l=1}^m C'_{i'_ls}$ and $C_{js}=C'_{js}$ for all $j\in N\backslash\{i\}$ and $s\in S$. Let us first see that $\sum_{l=1}^mR^U_{i'_l}(N',S,p,C')\geq R^U_i(N,S,p,C)$. We have that
        \begin{align*} 
        \sum_{l=1}^m R^U_{i'_l}(N',S,p,C')\geq R^U_i(N,S,p,C)\Leftrightarrow\frac{m\sigma}{|N|-1+m}\geq \frac{\sigma}{|N|}\Leftrightarrow |N|m\geq |N|+m-1\Leftrightarrow|N|\geq 1.
	\end{align*}

 Now, let us see that $\sum_{l=1}^mR^{SU}_{i'_l}(N',S,p,C')\geq R^{SU}_i(N,S,p,C)$. We have that
 \begin{align*} 
        \sum_{l=1}^m R^{SU}_{i'_l}(N',S,p,C')\geq R^U_i(N,S,p,C)\Leftrightarrow\sum_{l=1}^m\sum_{s\in S, i'_l\in N'_s}\frac{1}{|N'_s|}p_s\geq \sum_{s\in S, i\in N_s}\frac{1}{|N_s|}p_s,
	\end{align*}
 where $N_{s}=\{j\in N: C_{js}>0\}$ and $N'_{s}=\{j\in N': C'_{js}>0\}$ for all $s\in S$. Then it is sufficient to prove that if $i\in N_s$, then there exists at least one $i'_{l}\in N'_s$ for all $s\in S$. Since $C_{is}=\sum_{l=1}^m C'_{i'_ls}$ for all $s\in S$ if $C_{is}>0$ then $C'_{i'_ls}>0$ for some $l\in\{1,\hdots,m\}$, and the proof is concluded.
\end{proof}
\end{document}